\documentclass[12pt]{Thesis}
\begin{document}

\onehalfspacing
\pagestyle{fancy}
\pagenumbering{arabic}

\addtocounter{chapter}{2}

\begin{savequote}[4.0in]
  \vspace{6.0in}
  \large
  As far as the laws of mathematics refer to reality, they are not certain; and
  as far as they are certain, they do not refer to reality.
  \qauthor{Albert Einstein}
\end{savequote} 

\chapter[Smoothed Particle Hydrodynamics]{Smoothed Particle Hydrodynamics \\
\normalsize{Or: How I Learned to Stop Worrying and Love the Lagrangian*}}
\label{chapter3}
\small * With apologies to Drs. Strangelove and Price
\normalsize
\clearpage

\section{Introduction}
\label{c3intro}

The study of astrophysical phenomena presents a multitude of obstacles
to the potential student.  In addition to the usual obstacles of
understanding the physical properties of the system in question, the
sheer scale of astrophysical events renders laboratory experiments impossible
in the vast majority of cases.  To this end, it has been necessary to
assemble a new, numerical laboratory in the form of computational
simulations, and conduct experiments and analyses via this medium.
The growth of computing power over the past 80 years, from the Colossus of
Bletchley Park's Enigma code-cracking efforts in the 1940s, through ENIAC and
Los Alamos National Laboratory's modelling of thermonuclear detonations in the
1950s, up to the supercomputers of today, has in turn allowed the computational
domain to become a mainstay of astrophysical experimentation. 

Two principal approaches to computational simulations have evolved to
enable these numerical simulations.  Eulerian methods use geometric
grids, either fixed or adaptive (the so-called AMR or Adaptive Mesh
Refinement codes), with the fluid parameters evaluated over the grid
cells.  Such codes formed the basis of the revolution in Computational Fluid
Dynamics (CFD) that started in the late 1960s and early 70s, and as such they
remain the most widely used approach.  Applications of such codes cover a huge
range, from industrial aerodynamics in the automotive and aerospace sectors,
to stress calculations and solid mechanics for civil engineering and
architecture, to chemical reaction modelling and protein folding in
biomolecular models.    

Lagrangian methods on the other hand dispense with fixed points in space
and instead evolve the fluid equations in a co-moving frame.
A common approach is to use discrete particles that are carried with the flow
-- hydrodynamic (and other) properties are then evaluated at the particle
positions, and are calculated from a weighted average of the values on other
local particles.  In this manner each particle is essentially ``smoothed''
over a finite volume of fixed mass, and in this way these so-called Smoothed
Particle Hydrodynamics or SPH codes are naturally adaptive with density.
Although SPH was originally developed by the astrophysics community, it too
has found uses and applications in a much wider range of fields.  In
engineering it has been applied to dam breaks and atomised oil lubrication
flows, while a number of physics engines in computer games use SPH as a basis.
The community has grown to the point where there is now a Europe-wide network
of users called SPHERIC - the SPH European Research Interest
Community\footnote{http://wiki.manchester.ac.uk/spheric/index.php/SPHERIC\_Home\_Page}. This
aims to share advances in code development across the user community, and to
prevent the re-invention of the wheel when it comes the solution of known
problems. 

Each of these approaches has advantages and disadvantages with respect
to the other.  Generally speaking, AMR codes have a higher resolution
for a given number of grid cells than an SPH code with an equal number
of particles.  Furthermore, they can be made to adapt to any flow
parameter (although this is not always trivial!), while SPH adapts primarily
with density only.  On the other hand, SPH naturally handles vacuum boundary
conditions, whilst large grids are required with AMR codes to prevent the flow
disappearing from the edge of the computational domain.  As SPH is a
Lagrangian method, advection of flow properties is inherent, whereas this
presents problems for AMR codes, and which usually entails an unphysical
increase in entropy. In a similar manner, SPH codes can be implemented in such
a manner that they are inherently conservative of mass, momentum and energy,
and similarly, unless it is explicitly added in shocks, they likewise conserve
entropy.  Nonetheless it is emphatically \textbf{not} true to say that either
SPH or grid code methods are ``better'' than the other, simply that the more
appropriate approach should be chosen for any given problem, and indeed
greater confidence in the results will ensue if the two methods concur.

Having said that, throughout this chapter I shall however consider only the
SPH approach, as it is this one that I have used to generate all the results
discussed henceforth.  Furthermore, as all the problems I have considered have
been fully three-dimensional, throughout this chapter I shall consider only
the derivation and discussion of SPH in 3D.  This chapter is therefore
structured as follows:  In \sref{c3basics} I shall introduce the basic
concepts of the SPH method, then in \sref{c3fluids} this is used to re-write
the fluid equations in a manner that can be solved numerically.  Section
\ref{c3dissipation} discusses the dissipative processes required for the
correct implementation of artificial viscosity and the introduction of entropy
in shock waves, and then in \sref{c3smoothinglengths} I discuss how the SPH
formalism may be made more adaptive still by the self-consistent inclusion of
variable smoothing lengths.  As many astrophysical problems are strongly
influenced by gravitational forces I detail how these may be implemented in
\sref{c3gravity}.  In \sref{c3neighbours} I briefly summarise the methods used
to find the nearest neighbours, and then in \sref{c3timesteps} I consider how
the code is evolved forward in time, and various time-stepping
criteria. Finally in \sref{c3summary} I briefly outline the properties of the
code I have used, point the reader in the direction of some standard numerical
tests used for code evaluation and consider further extensions to the method.
\section{SPH Basics}
\label{c3basics}

In the following section I shall discuss the derivation of the
SPH formalism from first principles, showing how a continuous field
can be mapped on to (and thus approximated by) a series of discrete
particles, and the errors involved in this approximation.  I then show
how derivatives may be calculated, and discuss ways in which the
particles may be suitably smoothed to represent the field.

\subsection{Discrete Approximations to a Continuous Field}
\label{c3interpolant}

We start from the (mathematically) trivial identity
\begin{equation}
  f(\vec{r}) = \int \limits_{V} f(\vec{r}') \; \delta( \vec{r} -
  \vec{r}' ) \; \d \vec{r}',
\label{c3identity}
\end{equation}
where $f(\vec{r})$ is any (scalar) function defined on a three-dimensional
co-ordinate system $\vec{r}$ ranging over a volume $V$.  Similarly,
$\delta ( \vec{r} )$ is the Dirac delta function, and $\vec{r}'$
is a dummy variable also ranging over $V$.

We may generalise the delta function to a so-called smoothing kernel $W$ with
a characteristic width $h$ (known as the smoothing length) such that 
\begin{equation}
  \lim_{h \rightarrow 0}  W( \vec{r}, h ) = \delta ( \vec{r} ),
\label{c3kernellim}
\end{equation}
subject to the normalisation 
\begin{equation}
  \int \limits_{V} W( \vec{r}, h) \; \d \vec{r}' = 1.
\label{c3kernelnorm}
\end{equation}
By expanding $W(\vec{r} - \vec{r}',h)$ as a Taylor series, it can be shown
that for symmetric kernels $W(\vec{r} - \vec{r}',h) = W(\vec{r}' - \vec{r},
h)$, \eref{c3identity} becomes 
\begin{equation}
  f(\vec{r}) = \int \limits_{V} f(\vec{r}') \; W( \vec{r} -
  \vec{r}', h ) \; \d \vec{r}' + \ord{h^{2}},
\label{c3kernelapprox}
\end{equation}
the second order accuracy arising from the vanishing of the kernel gradient at
$\vec{r}' = \vec{r}$ (see for instance \citealt{Price05,Benz90,Monaghan92}).
Note that more elaborate kernels accurate to $\ord{h^{4}}$ can be constructed,
but these suffer from the problem that $W(\vec{r},h)$ can become negative in
certain ranges \citep{Price05,Monaghan92}, thus potentially leading to negative
density evaluations in certain pathological situations. 

Nonetheless for a second order, symmetric kernel, for any finite density $\rho
( \vec{r} )$ within $V$, \eref{c3kernelapprox} is exactly equivalent to
\begin{equation}
  f(\vec{r}) = \int \limits_{V} \frac{f(\vec{r}')}{\rho( \vec{r}')}
  \; W( \vec{r} - \vec{r}', h ) \; \rho ( \vec{r}')  \d
  \vec{r}' + \ord{h^{2}}. 
\end{equation}
Discretising this continuous field on to a series of particles
of (potentially variable) mass $m = \rho (\vec{r}') \d
\vec{r}'$, the original identity \eref{c3identity} becomes 
\begin{equation}
  f( \vec{r} ) \approx \sum_{i} \frac{m_{i}}{\rho_{i}} f(\vec{r}_{i}) \; W(
  \vec{r} - \vec{r}_{i}, h),
\label{c3discreteapprox}
\end{equation}
where now $f(\vec{r}_{i})$, $m_{i}$ and $\rho_{i} = \rho(\vec{r}_{i})$ are the
scalar value, mass and density of the $i^{th}$ particle, and $i$ ranges over
all particles within the smoothing kernel.  Equation \ref{c3discreteapprox}
therefore represents the discrete approximation to the continuous scalar field
$f$ at position $\vec{r}$ in the computational domain $V$, and is thus the
basis of all SPH formalisms.  Note that the position $\vec{r}$ at which the
function $f$ is approximated is completely general and is not restricted to
particle positions, although in practice this is where the values are actually 
evaluated. 

\subsection{Spatial Derivatives and Vector Calculus}
In order for the SPH discretisation of a field to be useful as a method of
solving fluid flows, it is clear that the spatial derivatives of any given
quantity must also have a suitable approximate form\footnote{Temporal
  derivatives are naturally also required, and these will be discussed in
  due course}.   Here therefore, I summarise the SPH approximations for various
vector calculus quantities.

\subsubsection{Gradient of a Scalar Field}
The approximation for the gradient of a scalar field can be derived by taking
the spatial derivative of \eref{c3identity}, and applying the smoothing
kernel.  Noting that $\nabla \equiv \partial / \partial \vec{r}$, we
therefore see that 
\begin{eqnarray}
  \nabla f ( \vec{r} ) & = & \frac{\partial}{\partial \vec{r}} \int
  \limits_{V} f(\vec{r}') \; \delta( \vec{r} - \vec{r}' ) \;
  \d\mathrm{r}' \\
  & = & \frac{\partial}{\partial \vec{r}} \int \limits_{V} f(\vec{r}')
  \; W( \vec{r} -  \vec{r}', h ) \;  \d\mathrm{r}' + \ord{h^{2}},
\end{eqnarray}
in a similar manner to \eref{c3kernelapprox}.  Given that the only part to
depend on $\vec{r}$ is the smoothing kernel $W$, and again introducing the
density $\rho(\vec{r}')$ in both the numerator and the denominator we
obtain
\begin{equation}
  \nabla f(\vec{r}) = \int \limits_{V}
  \frac{f(\vec{r}')}{\rho(\vec{r}')} \frac{\partial}{\partial
    \vec{r}} W(\vec{r} - \vec{r}', h) \rho(\vec{r}') \; \d
  \vec{r}' + \ord{h^{2}}.
\end{equation}
Finally this may be discretised in the same way as before, to give 
\begin{equation}
  \nabla f( \vec{r} ) \approx  \sum_{i} \frac{m_{i}}{\rho_{i}} f(\vec{r}_{i})
  \; \nabla W( \vec{r} - \vec{r}_{i}, h)
\label{c3discretegrad}
\end{equation}
as an estimator for the gradient of a scalar field $f(\vec{r})$.  Notable from
the above result is that the gradient of a scalar field can be approximated by
the values of the field itself along with the gradient of the kernel.
Computationally this is very useful as at no point does $\nabla f$ have to be
evaluated for any particle, whilst the gradient of the kernel will be known
explicitly for any sensible choice of $W$.  

\subsubsection{Divergence of a Vector Field}
Although \eref{c3identity} was given only for a scalar field, a similar
identity may be given for a vector field $\vec{F}(\vec{R})$, namely 
\begin{equation}
  \vec{F}(\vec{r}) = \int \limits_{V} \vec{F}(\vec{r}') \; \delta(
  \vec{r} - \vec{r}' ) \; \d \mathrm{r}',
\label{c3identityv}
\end{equation}
Taking the divergence of this with respect to $\vec{r}$, and noting once
again that the only term to depend on $\vec{r}$ is the smoothing kernel we
find that the integral approximation becomes 
\begin{equation}
  \nabla \cdot \vec{F}(\vec{r}) = \int \limits_{V}
  \vec{F}(\vec{r}') \cdot \nabla W(\vec{r} - \vec{r}',h)
  \d \vec{r}' + \ord{h^{2}},
\end{equation}
and thus as before this can be discretised to obtain the approximation
\begin{equation}
  \nabla \cdot \vec{F}(\vec{r}) \approx  \sum_{i} \frac{m_{i}}{\rho_{i}}
  \; \vec{F}(\vec{r}_{i}) \cdot \nabla W( \vec{r} - \vec{r}_{i},h). 
\label{c3discretediv}
\end{equation} 

\subsubsection{Curl of a Vector Field}
By a precisely similar argument, it is possible to show that the curl of a
vector $\vec{F}$, $\nabla \times \vec{F}$ can be approximated using
\begin{equation}
  \nabla \times \vec{F}(\vec{r}) \approx  \sum_{i} \frac{m_{i}}{\rho_{i}}
  \; \vec{F}(\vec{r}_{i}) \times \nabla W( \vec{r} - \vec{r}_{i},h), 
\label{c3discretecurl}
\end{equation}
although this is relatively little used unless magnetohydrodynamic (MHD)
effects are being taken into account.

\subsection{Errors}
The approximations given in
\erefsss{c3discreteapprox}{c3discretegrad}{c3discretediv}{c3discretecurl}
encompass both the $\ord{h^{2}}$ errors of considering only the integral
term, and also the errors inherent in the discretisation (which arise due to
incomplete sampling of the smoothing kernel).  In the former case we see that
the $\ord{h^{2}}$ errors are reduced by decreasing the smoothing length, while
the discretisation (sampling) errors are minimised by increasing the number of
particles within the smoothing kernel.  Barring numerical stability issues
\citep{Readetal09}, this discrete approximation is therefore at its most
accurate with large numbers of particles contained within a small smoothing
length.  However, this must be balanced against the need for computational
speed and efficiency, and hence there is a compromise to be struck. 

These errors are neatly illustrated by considering the approximations
to a constant function $f(\vec{r}) \equiv 1$ and the zero function,
which can be obtained by noting that with this definition of $f$, $\nabla
f(\vec{r}) = 0$.  The SPH approximations for one and zero therefore become 
\begin{eqnarray}
  1 & \approx & \sum_{i} \frac{m_{i}}{\rho_{i}} \; W(\vec{r} -
  \vec{r}_{i}, h), \\
\label{c3unitapprox}
  0 & \approx & \sum_{i} \frac{m_{i}}{\rho_{i}} \; \nabla W(\vec{r} -
  \vec{r}_{i}, h).
\label{c3zeroapprox}
\end{eqnarray}
Since in neither case does the equation reduce to an identity, we see that
there are inherent errors in estimating even constant functions.
Nonetheless, with suitable choices for the number of particles within
the smoothing kernel and the smoothing length, these may be kept to an
acceptable level.  For a more detailed derivation and discussion of
these errors, the reader is directed to \citet{Price05,Monaghan92,Benz90} and
\citet{Readetal09}.  

\subsection{Improved Approximations for Spatial Gradients}
Although the approximations given in
\erefsss{c3discreteapprox}{c3discretegrad}{c3discretediv}{c3discretecurl}  are
those that arise most readily from the SPH approximation, it is possible to
construct other estimators for the gradient of a scalar field.  For instance,
by noting that for any quantity $f(\vec{r}) \equiv 1.f(\vec{r})$, we see that  
\begin{equation}
  \nabla f(\vec{r}.1) = 1.\nabla f(\vec{r}) + f(\vec{r}) \nabla 1
\end{equation}
and therefore that
\begin{equation}
  \nabla f(\vec{r}) = \nabla f(\vec{r}) - f(\vec{r}) \nabla 1.
\label{c3gradA}
\end{equation}
Clearly, since $\nabla 1 = 0$ these forms should be identical.  From
\eref{c3unitapprox} however, we see that the SPH approximation for $\nabla 1$
is non-zero, and thus using \eref{c3gradA} we may define another estimate for
$\nabla f(\vec{r})$ as 
\begin{eqnarray}
  \nabla f(\vec{r}) & = & \sum_{i} \frac{m_{i}}{\rho_{i}}
  f(\vec{r}_{i}) \; \nabla W(\vec{r} -
  \vec{r}_{i}, h) - f(\vec{r}) \sum_{i} \frac{m_{i}}{\rho_{i}} \;
  \nabla W(\vec{r} - \vec{r}_{i}, h)\\
  & = & \sum_{i} m_{i} \frac{f(\vec{r}_{i}) - f(\vec{r})}{\rho_{i}} \; \nabla
  W(\vec{r} - \vec{r}_{i}, h).
\label{c3discretegrad0}
\end{eqnarray}
This approximation clearly has the advantage that it vanishes identically for
constant functions.

A more general class of interpolants arises from considering the vector
calculus identity 
\begin{equation}
  \nabla (f \rho^{n}) \equiv nf\rho^{n-1} \nabla \rho + \rho^{n}
  \nabla f,
\label{gradfrhon}
\end{equation}
valid for all $n \in \mathbb{R}$.  This in turn leads to the following
identity for $\nabla f$ 
\begin{equation}
  \nabla f \equiv  \frac{1}{\rho^{n}} \left[ \nabla (f\rho^{n}) - nf\rho^{n-1}
    \nabla \rho \right].
\end{equation}
Substituting $\rho$ and $f\rho^{n}$ into \eref{c3discretegrad}, we obtain a
general interpolant for $\nabla f$, such that 
\begin{equation}
  \nabla f(\vec{r}) = \frac{1}{\rho(\vec{r})^{n}} \sum_{i} m_{i} \left(
  f(\vec{r}_{i}) \rho(\vec{r}_{i})^{n-1} - nf(\vec{r}) \rho(\vec{r})^{n-1}
  \right) \; \nabla W(\vec{r} - \vec{r}_{i}, h).
\label{c3discretegradn}
\end{equation}
Two instances of this general case turn out to be particularly useful, namely
where $n = 1$ and $n = -1$.  For the former case we obtain
\begin{equation}
  \nabla f(\vec{r}) = \frac{1}{\rho(\vec{r})} \sum_{i} m_{i}
  \left( f(\vec{r}) - f(\vec{r}_{i}) \right) \nabla W(\vec{r} -
  \vec{r}_{i},h).
\label{c3discretegrad3}
\end{equation}
This is very similar in form to that given in \eref{c3discretegrad0}, with the
exception that knowledge of the density at $\vec{r}$ is required \emph{a
  priori}.  Although no longer anti-symmetric in $f(\vec{r})$ and $f_{i}$,
it is nonetheless exact for constant functions. 

In the case where $n = -1$ we obtain
\begin{equation}
  \nabla f(\vec{r}) = \rho(\vec{r}) \sum_{i} m_{i} \left(
  \frac{f(\vec{r})}{\rho(\vec{r})^{2}} +
  \frac{f(\vec{r}_{i})}{\rho(\vec{r}_{i})^{2}} \right) \; \nabla W(\vec{r} -
  \vec{r}_{i}, h). 
\end{equation}
While this form is no longer exact for constant functions, it is commonly used
as an estimator for the pressure gradient $(\nabla P) / \rho$, as it is
pairwise symmetric and as such ensures conservation of momentum.  This is also
the form of the gradient that arises naturally from a Lagrangian formulation
of the fluid equations, as I shall show in \sref{c3MomentumConservation}.

\subsection{Improved Divergence Estimates}
In a similar manner to the gradient, improved estimates can be made
for the divergence of a vector field.  By noting that
$\vec{F}(\vec{r}) = 1.\vec{F}(\vec{r})$, the estimate 
\begin{equation}
  \nabla \cdot \vec{F}(\vec{r}) \approx \sum_{i}
  \frac{m_{i}}{\rho_{i}} \left( \vec{F}(\vec{r}_{i}) -
    \vec{F}(\vec{r}) \right) \cdot \nabla W(\vec{r} -
  \vec{r}_{i}, h).
\end{equation}
can be arrived at, which again becomes exact for constant functions.
In a similar manner to the expansion given in \eref{gradfrhon}, a
general class of estimates can be arrived at by considering the
identity
\begin{equation}
  \nabla \cdot (\rho^{n} \vec{F}) = \rho^{n} \nabla \cdot
  \vec{F} + n \rho^{n-1} \vec{F} \cdot \nabla \rho,
\end{equation}
the $n = 1, -1$ cases of which are given by
\begin{equation}
  \nabla \cdot \vec{F}(\vec{r}) \approx
  \frac{1}{\rho(\vec{r})} \sum_{i} m_{i} \left( \vec{F}(\vec{r}_{i}) -
    \vec{F}(\vec{r}) \right) \cdot \nabla W(\vec{r} -
  \vec{r}_{i}, h)
\label{c3divn1}
\end{equation}
and
\begin{equation}
  \nabla \cdot \vec{F}(\vec{r}) \approx
  \rho(\vec{r}) \sum_{i} m_{i} \left(
  \frac{\vec{F}(\vec{r}_{i})}{\rho(\vec{r}_{i})^{2}} +
  \frac{\vec{F}(\vec{r})}{\rho (\vec{r})^{2}} \right) \cdot \nabla W(\vec{r} -
  \vec{r}_{i}, h)
\end{equation}
respectively.  Once again these estimates have the advantages of being
exact for constant functions in the former case and pairwise symmetric
in $(\nabla \cdot \vec{F})/\rho$ in the latter case.

\subsection{Smoothing Kernels}
From the above it is clear that the choice of smoothing kernel is an
important one.  It must by definition obey the criteria set out in
\erefs{c3kernellim}{c3kernelnorm} in that it must
tend to a $\delta$-function as $h \rightarrow 0$ and it must be
normalised so the area under the curve is unity.  For the purposes of
calculating the gradients of quantities it is also clear that it
should have a continuous and well defined first derivative, and from a
symmetry argument it should be spherically symmetric, and thus depend
only on $r = |\vec{r} - \vec{r}'|$ and $h$.

One of the first choices for the smoothing kernel was the Gaussian function,
such that  
\begin{equation}
  W(r,h) = \frac{1}{h^{3} \pi^{3/2}} e^{-x^{2}},  
\label{c1Gaussiankernel}
\end{equation}
where $x = r/h$.  However, this has the drawback that $W > 0$ for all $r$, and
thus all particles within the computational domain contribute.  The
computational cost of such a kernel therefore scales as $\ord{N^{2}}$,
where $N$ is the number of particles in the simulation.  Given that
(for purely hydrodynamical quantities) long range forces are
negligible, it makes sense to restrict the kernel to those with
compact support, i.e. make them subject to the condition that $W(r,h)
= 0$ where $r/h > k$ for some constant $k$.  This means that
the computational cost scales as $\ord{NN_{\mathrm{neigh}}}$, where
$N_{\mathrm{neigh}}$ is the average number of particles within a sphere of
radius $r = kh$ about any one particle.

For this reason, cubic spline kernels are often used (see \citealt{MonaghanL85}
for instance), where the kernel is defined as 
\begin{equation}
  W(r,h) = \frac{1}{\pi h^{3}} \left\{ 
    \begin{array}{ll} 
      \displaystyle 1 - \frac{3}{2} x^{2} + \frac{3}{4} x^{3} & 0 \leq x \le
      1; \\ 
      \displaystyle \frac{1}{4} \left(2 - x \right)^{3} & 1 \leq x \le 2; \\
      \displaystyle 0 & x \geq 2,
    \end{array}
  \right.
  \label{c3splinekernel}
\end{equation}
where $x = r/h$ as in \eref{c1Gaussiankernel}.  Here only particles
within $2h$ of the central particle contribute to the 
smoothing kernel, which is spherically symmetric and smoothly differentiable
for all $r$.  Although many other kernels are possible (see \citealt{Monaghan92,
  FulkQ96,Price05,Readetal09} for example) this is a commonly used kernel, and
is the one present in the code I have used throughout.     

Note from the above the gradient of the kernel is well defined for all
values of $x$, such that
\begin{eqnarray}
  \nabla W(r,h) 
  & = & \frac{\partial}{\partial r} W(r,h) \\
  & = & \frac{1}{\pi h^{4}} 
  \left\{
    \begin{array}{ll}
      \displaystyle \frac{9}{4}x^{2} - 3x  & 0 \leq x \le 1 \\
      \displaystyle -\frac{3}{4} (2-x)^{2} & 1 \leq x \le 2 \\
      \displaystyle 0                      & x \geq 2
    \end{array}
  \right. .
  \label{c3gradsplinekernel}
\end{eqnarray}  

Finally it is worth noting that in general the form of the kernel
makes little overall difference to the computational speed of the
code.  This is because most codes tabulate the values of both the kernel and
its gradient rather than compute them directly, and thus the
form of the kernel may be as simple or as complex as required, even
(theoretically at least) to the extent of being non-analytic functions.

\section{Fluid Equations}
\label{c3fluids}

Given that the SPH formalism has now been put on a sound mathematical footing,
in this section I shall use it to obtain approximations to the equations
governing fluid motion, such that they can be used to construct a viable
numerical algorithm for solving fluid flows.  For the dual purposes of brevity
and simplicity I shall here consider only the case of an inviscid compressible
flow in the absence of body forces, although the inclusion of both gravity and
(artificial) viscosity will be discussed in due course.  First however, it is
useful to summarise the principal equations of motion  in their standard
conservative form. 

The continuity (conservation of mass) equation is given by 
\begin{equation}
  \frac{\partial \rho}{\partial t} + \nabla \cdot (\rho \vec{v}) =
  0
\label{c3continuity}
\end{equation}
where as normal, $\rho$ is the density, $t$ is time and $\vec{v}$
is velocity.

The Euler equation gives the equations of motion in the case of an
inviscid fluid, and encapsulates the conservation of momentum.  In the
absence of external (body) forces it becomes
\begin{equation}
  \frac{\partial \rho \vec{v}}{\partial t} + \nabla \cdot (\rho \vec{v}
  \otimes \vec{v})  + \nabla P = \vec{0},
  \label{c3Euler}
\end{equation}
where $P$ is the fluid pressure and $\otimes$ represents the outer or tensor
product\footnote{The outer product of two vectors may be summarised as
  $\vec{A} \otimes \vec{B} = \vec{A}\vec{B}^{T} = A_{i}B_{j}$ (in indicial
  notation).}.  For compressible flows it is also necessary to take into account
the energy equation, and as such the conservation of energy is embodied in the
following equation;  
\begin{equation}
  \frac{\partial u}{\partial t} + \nabla \cdot
  \left[(u + P)\vec{v} \right] = 0,
\label{c3energy}
\end{equation}
where $u$ is the specific internal energy and $v = |\vec{v}|$ is the
magnitude of the velocity vector.  Finally it is worth noting that these five
equations (there are 3 components to the momentum equation) contain six
unknowns ($\rho$, the three components of velocity $v_{x}$, $v_{y}$ and
$v_{z}$, $P$ and $u$).  Therefore in order to solve the system we require a
further constraint; an equation of state is required.  All the analysis I
shall present henceforth uses the ideal gas equation of state, where 
\begin{eqnarray}
  P &=& \kappa (s) \rho^{\gamma}, \\
  \label{c3eosentropy}
  &=& (\gamma - 1) u \rho,
  \label{c3eos}
\end{eqnarray}
where $\gamma$ is the adiabatic index (the ratio of specific heats),
which throughout has been set to $5/3$, and $\kappa(s)$ is the
adiabat, itself a function of the specific entropy $s$.  In the case of
isentropic flows, $s$ and (thus $\kappa$) remains constant.

I shall now discuss the SPH formulation of each of the continuity,
momentum and energy equations in turn.  Note that again for the
purposes of brevity I assume that the smoothing length is held
constant (i.e. $\dot{h} = 0$, where the dot denotes the derivative
with respect to time), and is equal for all particles.  Individual, variable
smoothing lengths will be discussed in due course.  Furthermore, I assume
throughout that the mass of each particle is held constant, such that $m_{i} =
const$, and again that all particles are of equal mass.  Although it is
possible to have individually varying particle masses, the code I use does not
have this feature, and therefore I have not included a discussion of it here.
Finally note that from here onwards, all the approximations are evaluated at
specific particle positions, as this is how the SPH algorithm is implemented
within particle-based codes. 

\subsection{Conservation of Mass}
Using \eref{c3discreteapprox}, we see that in the case of the density,
the SPH approximation becomes very simple, namely that at particle $j$ the
density $\rho_{j}$ becomes
\begin{equation}
  \begin{split}
    \rho_{j} &= \sum_{i} m_{i} W(\vec{r}_{j} - \vec{r}_{i}, h), \\
             &= \sum_{i} m_{i} W_{ji},
  \end{split}
  \label{c3density}
\end{equation}
where we write $W_{ji} = W(\vec{r}_{j} - \vec{r}_{i}, h)$, and where by
symmetry, $W_{ji} = W_{ij}$.  Note that here and henceforth, as the SPH
formalism is a discrete approximation to the underlying continuous
medium, we assume equality between the estimator on the RHS and the SPH
quantity on the LHS. 

Taking the full time derivative of \eref{c3density} we obtain
\begin{equation}
  \frac{\d \rho_{j}}{\d t} = \sum_{i} m_{i} \left[ 
    \frac{\partial W_{ji}}{\partial \vec{r}_{j}} \cdot \frac{\d
      \vec{r}_{j}}{\d t} + 
    \frac{\partial W_{ji}}{\partial \vec{r}_{i}} \cdot \frac{\d
      \vec{r}_{i}}{\d t} + 
    \frac{\partial W_{ji}}{\partial h} \frac{\d h}{\d t} \right],
\end{equation}
and noting that 
\begin{equation*}
  \frac{\d \vec{r}_{j}}{\d t} = \vec{v}_{j}, \;\; \frac{\d \vec{r}_{i}}{\d t}
  = \vec{v}_{i}, \;\; \frac{\d h}{\d t} = 0,
\end{equation*}
we find that the time derivative of density becomes
\begin{equation}
  \begin{split}
    \frac{\d \rho_{j}}{\d t} 
    &= \sum_{i} m_{i} \; (\vec{v}_{j} \cdot \nabla_{j} W_{ji} +
    \vec{v}_{i} \cdot \nabla_{i} W_{ji})\\
    &= \sum_{i} m_{i} \; \vec{v}_{ji} \cdot \nabla_{j} W_{ji}
  \end{split}
  \label{c3drhodt}
\end{equation}
where we use $\vec{v}_{ji} = \vec{v}_{j} - \vec{v}_{i}$, and where we note
that the gradient of the kernel is antisymmetric, i.e. that
\begin{equation}
  \nabla_{i} W_{ji} = -\nabla_{j} W_{ji}.
  \label{c3gradkernel}
\end{equation}
From \eref{c3divn1}, we note that the RHS of \eref{c3drhodt} is simply an
estimator of $-\rho_{j} \nabla_{j} \cdot \vec{v}_{j}$.  Hence \eref{c3drhodt}
becomes 
\begin{equation}
  \frac{\d \rho_{j}}{\d t} = -\rho_{j} \nabla_{j} \cdot \vec{v}_{j},
\end{equation} 
which is simply a reformulation of the continuity equation
\eref{c3continuity} using the Lagrangian time derivative
\begin{equation}
  \frac{\d}{\d t} = \frac{\partial}{\partial t} + (\vec{v} \cdot \nabla),
  \label{c3lagrangianddt}
\end{equation}
in which the second term accounts for the advection of flow properties
through the fluid.  Therefore we see that the SPH estimate for density
\eref{c3density} is \emph{automatically} conservative of mass (as long as
\eref{c3divn1} is used as an estimate for the divergence of velocity). 

\subsection{Conservation of Momentum}
\label{c3MomentumConservation}
Although there are various ways of deriving the equations of motion
consistently with the SPH framework, a particularly appealing one is
to use the Lagrangian formalism.  As long as the discrete Lagrangian
functional preserves the fundamental symmetries of the underlying continuous
one, this confers the inherent advantages that the resulting SPH
equations of motion will automatically fulfil the requisite conservation laws
(through Noether's Theorem) and also that the only approximations made are in
the discretisation of the Lagrangian itself.

\subsubsection{Linear Momentum}
Defined as the total kinetic energy of the system minus the total
internal energy (for purely hydrodynamical flows), the Lagrangian functional
$\mathcal{L}$ for the fluid is 
\begin{equation}
  \mathcal{L(\vec{r},\vec{v})} = \int \limits_{V} \frac{1}{2}
  \rho \vec{v} \cdot \vec{v} - \rho u \; \d \vec{r},
\label{c3Lagrangian}
\end{equation}
where as before, $u$ is the specific internal energy.  For
later simplicity, we note that through the equation of state (\eref{c3eos})
the specific internal energy is a function of density and pressure
$u = u(\rho,P)$, which in turn are functions of
position.  This gives $ u = u(\vec{r})$.  Now if we
again make the discretisation $m_{i} = \rho \d \vec{r}$, the SPH estimate of the
Lagrangian becomes  
\begin{equation}
  \mathcal{L}(\vec{r}, \vec{v}) = \sum_{i} m_{i} \left(
    \frac{1}{2} \vec{v}_{i} \cdot \vec{v}_{i} -
    u_{i}(\vec{r}_{i}) \right),
\label{c3discreteLagrangian}
\end{equation} 
where $i$ ranges over all particles.

The equations of motion for particle $j$ are obtained from the
Lagrangian through the Euler-Lagrange equations, as follows;
\begin{equation}
  \frac{\d }{\d t} \left( \frac{\partial \mathcal{L}}{\partial
    \vec{v}_{j}} \right) - \frac{\partial \mathcal{L}}{\partial
    \vec{r}_{j}} = \vec{0}.
  \label{c3Leqsmotion}
\end{equation}
By considering each of the terms in this equation it is therefore possible
to obtain an SPH approximation to the equations of motion that remains
fully conservative.  If we therefore consider the derivative of the
Lagrangian with respect to the velocity at particle $j$, we find
\begin{eqnarray}
  \frac{\partial \mathcal{L}}{\partial \vec{v}_{j}} &=&
  \frac{\partial}{\partial \vec{v}_{j}} \sum_{i} m_{i} \left(
    \frac{1}{2} \vec{v}_{i} \cdot \vec{v}_{i} -
    u_{i}(\vec{r}_{i}) \right), \nonumber\\
  & = & m_{j} \vec{v}_{j},
  \label{c3dLdv}
\end{eqnarray}
noting that since the velocities are independent the differential is zero
unless $i = j$.

Considering now the second term in the Euler-Lagrange \eref{c3Leqsmotion} we
find that 
\begin{eqnarray}
  \frac{\partial \mathcal{L}}{\partial \vec{r}_j} = - \sum_{i} \left[
  \frac{\partial u_{i}}{\partial P_{i}} \frac{\partial
    P_{i}}{\partial \vec{r}_{j}} + \frac{\partial u_{i}}{\partial
    \rho_{i}} \frac{\partial \rho_{i}}{\partial \vec{r}_{j}} \right],
  \label{c3dLdr}
\end{eqnarray}
where we have used \eref{c3eos} to obtain the full derivative of the internal
energy.  In the isentropic (dissipationless) case we see that $\kappa(s)$ is
constant, and thus the pressure is a function of density only, leading to 
\begin{equation}
  \frac{\partial \mathcal{L}}{\partial \vec{r}_j} = - \sum_{i} \left[
    \frac{\partial u_{i}}{\partial P_{i}} \frac{\d
      P_{i}}{\d \rho_{i}} + \frac{\partial u_{i}}{\partial
      \rho_{i}} \right] \frac{\partial \rho_{i}}{\partial \vec{r}_{j}}.
\end{equation}
From the equation of state \eref{c3eos} we find that 
\begin{equation}
  \frac{\partial u_{i}}{\partial P_{i}} \frac{\d P_{i}}{\d
    \rho_{i}} + \frac{\partial u_{i}}{\partial \rho_{i}} =
  \frac{P_{i}}{\rho_{i}^{2}}, 
  \label{c3dUdrho}
\end{equation}
and thus the derivative of the Lagrangian with respect to the position of
particle $j$ becomes
\begin{equation}
  \frac{\partial \mathcal{L}}{\partial \vec{r}_j} =
  - \sum_{i} \frac{P_{i}}{\rho_{i}^{2}} \frac{\partial\rho_{i}}{\partial
    \vec{r}_{j}}. 
  \label{c3dLdr1}
\end{equation}
Using \eref{c3density} we find that
\begin{eqnarray}
  \frac{\partial \rho_{i}}{\partial \vec{r}_{j}} 
  &=& \sum_{k} m_{i} \frac{\partial W_{ik}}{\partial \vec{r}_{j}} \\
  &=& \sum_{k} m_{i} \frac{\partial W_{ik}}{\partial r_{ik}} \frac{\partial
    r_{ik}}{\partial \vec{r}_{j}},
  \label{c3drhodr1}
\end{eqnarray}
where we take $r_{ik} = |\vec{r}_{ik}|$, and use the fact that the kernel is
spherically symmetric.  By direct differentiation,
\begin{equation}
  \frac{\partial r_{ik}}{\partial \vec{r}_{j}} = (\delta_{ij} - \delta_{kj})
  \hat{\vec{r}}_{ik}, 
\end{equation}
with $\hat{\vec{r}}_{ik} = \vec{r}_{ik} / r_{ik}$ the unit vector in the
direction of $\vec{r}_{ik}$.  Substituting this back into \eref{c3drhodr1} we
find that 
\begin{eqnarray}
  \frac{\partial \rho_{i}}{\partial \vec{r}_{j}} 
  &=& \sum_{k} m_{k} \frac{\partial W_{ik}}{\partial r_{ik}} (\delta_{ij} -
  \delta_{kj}) \hat{\vec{r}}_{ik} \\
  &=& \sum_{k} m_{k} \nabla_{j} W_{ik} \; \left(\delta_{ij} - \delta_{kj}
  \right),
  \label{c3drhodr}
\end{eqnarray}
where in the second case we have used the fact that $\partial / \partial
\vec{r}_{j} \equiv \nabla_{j}$.  

With reference to \eref{c3dLdr1} we find therefore that 
\begin{eqnarray}
  \frac{\partial \mathcal{L}}{\partial \vec{r}_{j}} &=&
  - \sum_{i} m_{i} \frac{P_{i}}{\rho_{i}^{2}} \sum_{k} m_{k} \nabla_{j}
  W_{ik} \left( \delta_{ij} - \delta_{kj} \right) \\
  & = & - m_{j} \frac{P_{j}}{\rho_{j}^{2}} \sum_{k} m_{k}  \nabla_{j} W_{jk} -
  \sum_{i} m_{i} m_{j} \frac{P_{i}}{\rho_{i}^{2}} \nabla_{j} W_{ij} \\
  & = & - m_{j} \sum_{i} m_{i} \left( \frac{P_{j}}{\rho_{j}^{2}} +
    \frac{P_{i}}{\rho_{i}^{2}} \right) \nabla_{j} W_{ji},
  \label{c3dLdrj}
\end{eqnarray}
where we have changed the summation index in the first term to $i$ and used
the fact that the gradient of the kernel is antisymmetric, i.e. that
$\nabla_{j} W_{kj} = -\nabla_{j} W_{jk}$.  Finally, by substituting
\erefs{c3dLdv}{c3dLdrj} into \eref{c3Leqsmotion} and dividing through by the
common factor $m_{j}$, we find that the SPH equations of motion become
\begin{equation}
  \frac{\d \vec{v}_{j}}{\d t} = - \sum_{i} m_{i} \left(
    \frac{P_{j}}{\rho_{j}^{2}} + \frac{P_{i}}{\rho_{i}^{2}} \right)
    \nabla_{j} W_{ji}.
    \label{c3eqnmotion}
\end{equation}
Since this equation is pairwise symmetric in $i,j$, it is clear that
the pressure force on particle $j$ due to particle $i$ is equal and
opposite (due to the antisymmetry of the kernel gradient) to the force
on particle $i$ from particle $j$.  In this manner, it is clear that
this formulation of the equation of motion conserves linear momentum
by construction.

\subsubsection{Angular Momentum}
To check that angular momentum $\vec{L} = \vec{r} \times m\vec{v}$ is
conserved, we note that its derivative with respect to time should be zero.
By using \eref{c3eqnmotion} we see that the time derivative of the angular
momentum of particle $j$ is given by 
\begin{eqnarray}
  \frac{\d \vec{L}_{j}}{\d t} & = &
  m_{j} \vec{v}_{j} \times \vec{v}_{j} + m_{j} \vec{r}_{j} \times
  \frac{\d \vec{v}_{j}}{\d t} \\
  & = & - m_{j} \sum_{i} m_{i} \left( \frac{P_{j}}{\rho_{j}^{2}} +
    \frac{P_{i}}{\rho_{i}^{2}} \right) \vec{r}_{j} \times \nabla_{i}
  W_{ij},
\end{eqnarray}
since by definition $\vec{v}_{j} \times \vec{v}_{j} = \vec{0}$. The total time
derivative of the angular momentum is therefore given by the sum over all
particles $j$, such that  
\begin{equation}
  \frac{\d \vec{L}}{\d t} = - \sum_{j} \sum_{i} m_{j} m_{i}
  \left( \frac{P_{j}}{\rho_{j}^{2}} + \frac{P_{i}}{\rho_{i}^{2}} \right)
  \vec{r}_{j} \times \nabla_{i} W_{ij}.
\end{equation}
Hence we see that by reversing the summation indices the entire
sum is \emph{anti}symmetric in $i$ and $j$, i.e.
\begin{eqnarray}
  \frac{\d \vec{L}}{\d t} 
  &=& -\sum_{j} \sum_{i} m_{j} m_{i} \left( \frac{P_{j}}{\rho_{j}^{2}} +
  \frac{P_{i}}{\rho_{i}^{2}} \right) \vec{r}_{j} \times \nabla_{i} W_{ij}, \\
  &=&  \sum_{i} \sum_{j} m_{i} m_{j} \left( \frac{P_{i}}{\rho_{i}^{2}} +
  \frac{P_{j}}{\rho_{j}^{2}} \right) \vec{r}_{i} \times \nabla_{j} W_{ji},
\end{eqnarray}
which can only be the case where the total sum is zero.  Hence the
angular momentum is constant with time, and thus angular momentum is
explicitly conserved.

\subsection{Conservation of Energy}
\label{c3energyconservation}
In the case of a purely hydrodynamical flow, the total energy $E = \rho u +
\rho v^{2} / 2$ is given by the sum of the kinetic and internal energies, such
that the SPH estimator becomes 
\begin{equation}
  E = \sum_{i} m_{i} \left( \frac{1}{2} \vec{v}_{i} \cdot \vec{v}_{i}
    + u_{i} \right).
\end{equation}
Clearly, where energy is conserved the time derivative of the total
energy should be zero. Taking the time derivative therefore, we find
that
\begin{eqnarray}
  \frac{\d E}{\d t} &=& \sum_{i} m_{i} \left( \vec{v}_{i} \cdot \frac{\d
      \vec{v}_{i}}{\d t} + \frac{\partial u_{i}}{\partial P_{i}}
  \frac{\d P_{i}}{\d t} + \frac{\partial u_{i}}{\partial \rho_{i}}
  \frac{\d \rho_{i}}{\d t} \right),
  \label{c3dEdt0} \\
  &=& \sum_{i} m_{i} \left( \vec{v}_{i} \cdot \frac{\d\vec{v}_{i}}{\d t} +
  \frac{P_{i}}{\rho_{i}^{2}} \; \frac{\d \rho_{i}}{\d t} \right),
  \label{c3dEdt01}
\end{eqnarray}
where we have again used the fact that in the dissipationless case $P =
P(\rho)$ and we can therefore amalgamate the latter two terms of the RHS of
\eref{c3dEdt0} using \eref{c3dUdrho}.  Using also the
equation of motion derived above in \eref{c3eqnmotion} and $\d \rho /\d t$
from the continuity equation \ref{c3drhodt} we therefore find that 
\begin{eqnarray}
  \frac{\d E}{\d t} &=& - \sum_{i} m_{i} \vec{v}_{i} \cdot
    \sum_{j} m_{j} \left( \frac{P_{i}}{\rho_{i}^{2}} +
    \frac{P_{j}}{\rho_{j}^{2}} \right) \nabla_{j} W_{ji} \nonumber \\
    && \qquad 
    +  \sum_{i} m_{i} \frac{P_{i}}{\rho_{i}^{2}} \; \sum_{j} m_{j}
    (\vec{v}_{i} - \vec{v}_{j}) \cdot \nabla_{j} W_{ji}  \\ 
  \label{c3dEdt1}
  & = & \sum_{i} \sum_{j} m_{i} m_{j} \left( \frac{P_{j}}{\rho_{j}^{2}}
  \vec{v}_{i} + \frac{P_{i}}{\rho_{i}^{2}} \vec{v}_{j} \right) \cdot
  \nabla_{j} W_{ji}, 
  \label{c3dEdt2}
\end{eqnarray}
where we have again used the fact that the kernel is antisymmetric to obtain
\eref{c3dEdt2}. Now using the same argument we used to show that angular
momentum is conserved, we note that \eref{c3dEdt2} is antisymmetric
under a reversal of $i$ and $j$, and thus must be equal to zero.  Hence we
find that 
\begin{equation}
  \frac{\d E}{\d t} = 0,
\end{equation}
and therefore that the total energy is also explicitly conserved.  

A corollary of this is that the time derivative of the internal energy is
given by the second term on the RHS of \eref{c3dEdt01}, such that 
\begin{eqnarray}
  \frac{\d u_{j}}{\d t} 
  &=& \frac{P_{j}}{\rho_{j}^{2}} \; \frac{\d \rho_{j}}{\d t},
  \label{c3dUdtdef} \\
  &=& \frac{P_{j}}{\rho_{j}^{2}} \; \sum_{i} m_{i} \; (\vec{v}_{j} -
  \vec{v}_{i}) \cdot \nabla_{i} W_{ji},
  \label{c3dUdt}
\end{eqnarray}
and indeed this is how the internal energy is evolved within SPH codes. 

It is worth noting that the formulation of SPH outlined above is therefore
explicitly conservative of mass, momentum (in both forms) and energy.  Hence,
while there are inevitably errors inherent in the SPH discretisation of a
continuous medium, these are the \emph{only} errors that appear, at least in
the case of a dissipationless hydrodynamical flow.

\section{Dissipative Effects}
\label{c3dissipation}

So far we have assumed the fluid flow to be barotropic (i.e.  $P = P(\rho)$),
and polytropic, with the polytropic index set equal to the adiabatic index
$\gamma$, the ratio of specific heats.  This in turn means that the flow is
isentropic, and therefore completely dissipationless.  While this is an
adequate approximation for many incompressible, inviscid and unshocked
compressible flows, it presents serious problems when it comes to modelling
transonic and supersonic flow regimes, as the conversion of mechanical
(kinetic) energy into heat (internal) energy is not correctly captured.  The
problem occurs because at a shock front, flow properties such as the velocity,
pressure, density and entropy change very rapidly, on the order of the mean
free path of the gas particles.  On large scales therefore these changes
appear discontinuous, and flow solvers that do not resolve the mean free path
(which is all of them) break down due to the apparently singular flow
gradients.   

There are two principal workarounds that allow numerical codes to solve
shocked flows.  One is to use a Riemann solver in a Gudonov-type code (see for
instance \citealt{Inutsuka02,ChaWhitworth03}), but I shall not go into any
detail here as this is not the approach used in the code I have used.  The
alternative approach, used in the majority of SPH codes, is to broaden the
shock across a small number of smoothing lengths.  This ensures that the flow
gradients do \emph{not} become infinite, and gives the correct asymptotic
behaviour away from the shock. This latter method is implemented by including
an \emph{artificial} dissipative term in the momentum and energy equations
that is triggered only in the presence of shocks, and it is this method that I
shall consider here.  

\subsection{Standard Artificial Viscosity Prescription}
Due to the fact that by construction, shock capturing through a viscous
process is an artificial one, there is considerable latitude in the way in
which such an artificial viscosity may be implemented.  This being said, it
must obey the following general rules \citep{vonNeumannRichtmyer50,Rosswog09}:
\begin{itemize}
  \item[-] The flow equations should contain no discontinuities;
  \item[-] The shock front should be of the order of a few times the smoothing
    length;
  \item[-] The artificial viscosity should reduce to zero away from the shock
    front; 
  \item[-] The Rankine-Hugoniot equations should hold over length scales
    larger than that over which the shock is smoothed, i.e. 
    \begin{eqnarray}
      \rho_{0} v_{0} &=& \rho_{1} v_{1}, \\
      P_{0} + \frac{\rho_{0} v_{0}^{2}}{2}  &=& P_{1} + \frac{\rho_{1}
        v_{1}^{2}}{2,} \\
      \frac{P_{0}}{\rho_{0}} + u_{0} + \frac{v_{0}^{2}}{2} &=& 
      \frac{P_{1}}{\rho_{1}} + u_{1} + \frac{v_{1}^{2}}{2},
    \end{eqnarray}
    where the subscripts 0 and 1 refer to pre- and post-shock regions
    respectively. 
  \item[-] The overall conservation of momentum and energy should not be
    adversely affected, while the entropy should rise from the pre- to
    post-shock regions.
\end{itemize}
By considering the SPH approximation to the momentum \eref{c3eqnmotion} where
the force is based on pairwise addition of terms of the form $P/\rho^2$, on
dimensional grounds it seems sensible to consider an artificial viscosity term
$\Pi$ of the form   
\begin{equation}
  \Pi \propto \frac{v^{2}}{\rho}
\end{equation}
for some suitable velocity scale $v$.  \citet{vonNeumannRichtmyer50} suggested
a viscous term dependent on the squared velocity divergence (which gives an
indication of the local expansion or contraction of the fluid), which
translates into SPH form as 
\begin{equation}
  (\Pi_{ij})_{\mathrm{NR}} = \frac{\beta_{\mathrm{SPH}} \, h^{2} |\nabla \cdot
    \vec{v}_{ij}|^{2}}{\bar{\rho}_{ij}},
\end{equation}
where $h$ represents a characteristic length scale (in SPH this is equivalent
to the smoothing length), $\bar{\rho}_{ij}$ is the average density of
particles $i$ and $j$ and $\beta_{\mathrm{SPH}}$ is a constant term of order
unity.  Noting that to first order  
\begin{eqnarray}
  |\nabla \cdot \vec{v}_{ij}| &=& \frac{|\vec{v}_{ij}|}{|\vec{r}_{ij}|}\\
  &\approx& \frac{|\vec{v}_{ij} \cdot \vec{r}_{ij}|}{|\vec{r}_{ij}|^{2} +
    \epsilon h^{2}} 
\end{eqnarray}
where as previously $\vec{r}_{ij} = \vec{r}_{i} - \vec{r}_{j}$ and where we
have added the extra (small) term in the denominator to prevent it becoming
singular, this von Neumann-Richtmyer term becomes 
\begin{equation}
  (\Pi_{ij})_{\mathrm{NR}} = \frac{\beta_{\mathrm{SPH}}\,
    \mu_{ij}^{2}}{\bar{\rho}_{ij}}
\end{equation}
where
\begin{equation}
  \mu_{ij} = \frac{h \vec{v}_{ij} \cdot \vec{r}_{ij}}{|\vec{r}_{ij}|^{2} +
    \epsilon h^{2}}.  
  \label{c3mu}
\end{equation}
By considering the bulk and shear viscosities of a generalised fluid it is
possible to obtain a second form of the artificial viscosity, and indeed this
has been known for some time \citep{Landshoff30,LandauLifshitz59}.  This form
is linear in the velocity divergence and uses the average sound
speed\footnote{Where as usual, the sound speed is defined as $\cs^{2} = \d P /
  \d \rho$.} $\bar{c}_{\mathrm{s},ij}$ as a second, characteristic velocity
component, giving the overall form  
\begin{equation}
  (\Pi_{ij})_{\mathrm{b}} = -\frac{\alpha_{\mathrm{SPH}} \,
    \bar{c}_{\mathrm{s},ij} \, \mu_{ij}}{\bar{\rho}_{ij}}, 
\end{equation}
where $\mu_{ij}$ is as before, and $\alpha_{\mathrm{SPH}}$ is a second
constant of order unity.  Note that the negative on the RHS arises from the
requirements that the viscous force component must be non-negative
(i.e. $\Pi_{ij} > 0$) and that it should be present only for convergent
flows, where $\vec{v}_{ij} \cdot \vec{r}_{ij} < 0$.  In fact these criteria
also hold for the von Neumann-Richtmyer form of the viscosity, and therefore
in both cases the viscosity is set to zero in expanding flow conditions. 

These two forms have different and complementary numerical effects.  At low
Mach numbers ($\mathcal{M} \lesssim 5$) the linear form performs very well in
shock tube tests \citep{Monaghan85}, whereas for stronger shocks it fails to
prevent inter-particle penetration \citep{Lattanzioetal85}.  This is
an unphysical phenomenon in which the two streams pass through each other at the
shock front, leading to the possibility of two particles occupying the
same position with differing velocities -- a multi-valued velocity
field.  This possibility can be prevented by using the quadratic form
of \citeauthor{vonNeumannRichtmyer50} as it provides a stronger
viscosity for high Mach number, although conversely, on its own this
decays too rapidly at low Mach numbers and fails to damp out the
unphysical post-shock oscillations or ``ringing'' that occurs.  The standard
solution is therefore to use the sum of the two terms \citep{Monaghan89},
resulting in a ``standard'' SPH viscous term of the form 
\begin{equation}
  \Pi_{ij} = \left\{ 
  \begin{array}{ll}
    \displaystyle \frac{-\alpha_{\mathrm{SPH}} \, c_{\mathrm{s},ij} \mu_{ij} +
    \beta_{\mathrm{SPH}} \, \mu_{ij}^{2}}{\rho_{ij}} \qquad &  \vec{v}_{ij}
    \cdot \vec{r}_{ij} < 0, \\ 
    \displaystyle 0 & \mbox{ otherwise. }
  \end{array}
  \right.
  \label{c3artvisc}
\end{equation}
Various numerical tests have showed that in general the constant values
$\alpha_{\mathrm{SPH}} = 1$, $\beta_{\mathrm{SPH}} = 2$  and $\epsilon  =
0.01$ in \eref{c3mu} give good results without significantly affecting
non-shocked flows. However, throughout the simulations discussed in the later
chapters of this thesis we have used values of $\alpha_{\mathrm{SPH}} = 0.1$
and $\beta_{\mathrm{SPH}} = 0.2$, which have been found to be adequate to
accurately resolve (weak) shocks, while at the same time minimising the
artificial heating which would have biased our simulation results -- details
can be found in \citet{LodatoR04}.

This general form of the viscosity can then be incorporated into the momentum
equation to give the following form;
\begin{equation}
  \frac{\d \vec{v}_{j}}{\d t} = - \sum_{i} m_{i} \left(
  \frac{P_{j}}{\rho_{j}^{2}} + \frac{P_{i}}{\rho_{i}^{2}} + \Pi_{ji} \right)
  \nabla_{j} W_{ji}.
  \label{c3disseqnmotion}
\end{equation}
Given that the artificial viscosity term is also pairwise symmetric in $i,j$
(since both $\vec{r}_{ij}$ and $\vec{v}_{ij}$ are anti-symmetric in $i$ and
$j$) it is clear that this form of the equation of motion also conserves
momentum exactly.  Likewise it is clear that angular momentum is conserved,
and furthermore, by a similar argument to that presented in
\sref{c3energyconservation} it is possible to show that in order to preserve
energy conservation, the energy equation must be modified to include an extra
dissipative term such that  
\begin{equation}
  \frac{\d u_{j}}{\d t} = \frac{P_{j}}{\rho_{j}^{2}} \; \sum_{i}
  m_{i} \vec{v}_{ji} \cdot \nabla_{i} W_{ji} + \sum_{i} m_{i}
  \Pi_{ji} \vec{v}_{ji} \cdot \nabla_{i} W_{ji}.
  \label{c3dissdUdt}
\end{equation}
In this manner it is therefore possible to include a dissipative term such
that shocks can be accurately captured, albeit broadened across a few smoothing
lengths.  Since mass, momentum and energy are still explicitly conserved
across the shock the Rankine-Hugoniot equations are automatically satisfied
at distances greater than a few smoothing lengths from the shock.
Furthermore, since (in theory at least) the artificial viscosity is zero away
from shocks, all the other initial criteria are satisfied also.  However,
there are various improvements that can be implemented, and these will now be
briefly discussed.  

\subsection{More Advanced Viscosities}
The thorn in the side of all viscosity prescriptions is the requirement that
in the absence of shocks or other natural dissipative processes the
artificial viscosity should reduce to zero, thereby requiring some means to
discriminate between shocks and other flow features.  Compounding the problem
is the fact that careful consideration of the artificial viscosity given above
(see for instance \citealt{LodatoPrice10}) shows that it provides both a bulk
and a shear viscosity, while to resolve shocks only the bulk component is
required.  Any artificial viscosity in the form of \eref{c3artvisc} therefore
necessarily introduces an unrequired shear viscosity, which can be problematic
in situations where shear flows are important (such as discs), leading to
spurious energy and angular momentum transport.   Furthermore, since
the shear force across any given particle varies with its smoothing
length, it is clear that this shear component is resolution dependent.
Generally speaking however this effect can be reduced by sensible
choices for $\alpha_{\mathrm{SPH}}$ and $\beta_{\mathrm{SPH}}$
\citep{LodatoR04} -- this further explains the low values of
$\alpha_{\mathrm{SPH}}$ and $\beta_{\mathrm{SPH}}$ mentioned earlier. 

\subsubsection{The Balsara Switch}
An attempt to reduce the induced viscosity in shear flows was presented by
\citet{Balsara95}, in which the standard artificial viscosity term $\Pi_{ij}$
is diminished by the factor $f_{ij} = |f_{i} + f_{j}| / 2$, where 
\begin{equation}
  f_{i} = \frac{|\nabla \cdot \vec{v}_{i}|}{|\nabla \cdot \vec{v}_{i}| +
    |\nabla \times \vec{v}_{i}| + 0.0001 c_{\mathrm{s},i} / h}.
  \label{c3Balsara}
\end{equation}
The inclusion of the vorticity (the curl of the flow field) allows this form
of the viscosity to perform better in shearing and obliquely shocked flows (see
for instance \citealt{Steinmetz96}), while remaining unaffected in the case of
normal shocks.  In a similar manner to the ``standard'' artificial viscosity
term, this form also includes a small term $0.0001 c_{\mathrm{s},i} / h$ to
prevent the viscosity from becoming singular.

\subsubsection{The Morris \& Monaghan Switch}
Although the Balsara switch represents a considerable improvement over the
standard form of artificial viscosity, problems still arise in the case of
shocks in shearing flows, such as those found in accretion discs.  For this
reason, \citet{MorrisMonaghan97} introduced the idea of a time-variant
viscosity such that $\Pi_{ij}$ remains unchanged from the standard form, but
where $\alpha_{\mathrm{SPH}} = \alpha(t)$, and where $\beta_{\mathrm{SPH}} =
2 \alpha_{\mathrm{SPH}}$.  The value of $\alpha$ is then evolved for each
particle according to the following equation;    
\begin{equation} 
  \frac{\d \alpha}{\d t} = - \frac{\alpha - \alpha_{\mathrm{min}}}{\tau} +
  S_{\vec{v}}.
\end{equation}
Here $\alpha_{\mathrm{min}} \sim 0.1 $ is some minimum value, justified by the
requirement that \emph{some} level of artificial viscosity is required to
maintain particle order\footnote{Note that only very low levels of viscosity
  are required for this purpose -- $\alpha_{\mathrm{min}} \sim 0.01$ should
  suffice \citep[in prep.]{CullenDehnen10}.}, $\tau \sim 0.1
- 0.2 \; h/\cs$ is a decay timescale (chosen so that the viscosity decays away
over a few smoothing lengths) and $S_{\vec{v}} = \max(-\nabla \cdot \vec{v},
0)$ is a source term, activated whenever the flow becomes convergent.
Although this form of the source term is still non-zero for pure shear flows,
this is counter-balanced to some extent by the decay term, and has been found
to work well in many tests of the artificial viscosity \citep{Dolagetal05}.
Further variations on this theme have been effected, including incorporating
the Balsara switch into the $\Pi_{ij}$ term, and capping the maximum value to
which $\alpha$ can rise by using a source term of the form $S_{\vec{v}} =
\max((\alpha_{\mathrm{max}} - \alpha) \nabla \cdot \vec{v},0)$
\citep{Rosswogetal00}.  For a good general overview of the relative merits of
a variety of artificial viscosity methods, see for instance
\citet[in prep.]{Lombardietal99,Rosswog09,CullenDehnen10}. 

\subsection{A Note on Entropy}
All of the above methods have essentially been aiming to capture the same
phenomenon, namely the increase in entropy found across a shock front, while
simultaneously ensuring isentropic flow elsewhere.  Furthermore all share the
common feature that flow evolution proceeds via integration of the energy
equation.  However, an approach espoused by \citet{SpringelH02} is to consider
evolving the entropy directly, thereby ensuring that the entropy can only
\emph{increase}.

In this manner, we recall that in terms of density $\rho$ and specific
entropy $s$, the equation of state is given by
\begin{equation}
  P_{i} = \kappa_{i}(s_{i}) \rho_{i}^{\gamma}.
\end{equation}
for some entropic function $\kappa(s)$.  Similarly, the internal energy
$u_{i}$ may be obtained from $\rho$ and $s$ via
\begin{equation}
  u_{i} = \frac{\kappa_{i}(s_{i})}{\gamma - 1} \rho_{i}^{\gamma - 1}.
\end{equation}
In the case of isentropic flow, we have $\kappa(s) = const$, and thus by
definition
\begin{equation}
  \frac{\d \kappa_{i}}{\d t} = 0.
\end{equation}
However, in the case where artificial viscosity is included, the time
derivative of the entropic function becomes
\begin{equation}
  \frac{\d \kappa_{i}}{\d t} = \frac{1}{2} \frac{\gamma - 1}{\rho_{i}^{\gamma
      - 1}} \; \sum_{j} m_{j} \Pi_{ij} \vec{v}_{ij} \cdot \nabla_{i} W_{ij}.
  \label{c3entropydot}
\end{equation}
By noting that 
\begin{equation}
  \nabla_{i} W_{ij} = |\nabla_{i} W_{ij}| \hat{\vec{r}}_{ij},
\end{equation}
and also that $\Pi_{ij}$ is only non-zero for $\vec{v}_{ij} \cdot \vec{r}_{ij}
< 0$, it is clear that the term on the RHS of \eref{c3entropydot} is strictly
non-negative, and thus that entropy can only increase throughout the flow.
Using this method of evolving the flow properties it is therefore possible to
explicitly ensure that the entropy of any particle increases monotonically
with time.  

\section{Variable Smoothing Lengths}
\label{c3smoothinglengths}

Up to now, it has been assumed that the smoothing length $h$ is held constant
with time, and is moreover equal for all particles.  In regions where
the density (and thus the number of neighbours) is roughly constant,
this maintains a constant (small) sampling error within the SPH
smoothing kernel. This requirement of constant smoothing length is quite
restrictive however, as it prevents the code adapting effectively to regions
of higher or lower than average density \citep{SteinmetzMuller93}.  By allowing
the smoothing length to vary both temporally and spatially, sampling
errors can be minimised across regions of varying density, as either
the number of neighbours or the mass within a smoothing kernel (and
thus the resolution) may be maintained. There are various simple ways
of allowing variable effective smoothing lengths that have been introduced, for 
instance \citet{Benz90} suggested using a symmetrised smoothing length $h_{ij}
= (h_{i} + h_{j})/2$, such that the kernel becomes
\begin{equation}
  W_{ij} = W\left( \vec{r}_{ij}, \frac{h_{i} + h_{j}}{2} \right).
\end{equation}
An alternative method has been suggested by \citet{HernquistKatz89}, in which
the average kernel value is used rather than the average smoothing length,
such that
\begin{equation}
  W_{ij} = \frac{W( \vec{r}_{ij},h_{i}) +  W(\vec{r}_{ij},h_{j})}{2}.
  \label{c3avkernel}
\end{equation}
With variable smoothing lengths it then becomes necessary to determine the
value of $h$ for each particle.  A standard method of doing this is to link
the smoothing length to the local density, such that 
\begin{equation}
  \rho_{i} h_{i}^{3} = const.
\end{equation}
Since this constant clearly has units of mass, it is frequently linked to the
particle mass, giving the following prediction for the particle smoothing
length;
\begin{equation}
  h_{i} = \eta \left( \frac{m_{i}}{\rho_{i}}\right)^{1/3},
  \label{c3hrho}
\end{equation}
where the coupling constant is generally in the range $1.2 < \eta < 1.5$
\citep{Rosswog09}.  By construction this method maintains a constant mass
within the smoothing kernel.  As each of the above formalisms remains pairwise
symmetric, momentum remains fully conserved, and increased spatial resolution is
achieved at relatively low computational cost.  The latter method
(using the averaged kernel value as in \eref{c3avkernel}) has
additional advantages in it is less problematic across shocks, and it couples
better with tree methods for calculating self-gravity
\citep{SteinmetzMuller93}.  Nonetheless, in both cases errors appear in either
the entropy or energy equation, such that either 
\begin{equation}
  \frac{\d E}{\d t} \mbox { or } \frac{\d \kappa(s)}{\d t} \sim \frac{\partial
    W}{\partial h} \frac{\partial h}{\partial t} \neq 0,
\end{equation}
\citep{Hernquist93}, and the relevant quantity is therefore not explicitly
conserved. 

It is however possible to construct SPH estimates that self-consistently
account for the variation in smoothing length, and therefore ensure
exact energy conservation.  In this case, the estimator
for density \eref{c3density} becomes
\begin{equation}
  \rho_{j} = \sum_{i} m_{i} W(\vec{r}_{ji}, h_{j}),
  \label{c3densitygradh}
\end{equation}
noting that the smoothing length used in the kernel is that associated with
particle $j$ only, and thereby remains constant throughout the summation.
By taking the (Lagrangian) time derivative, we obtain 
\begin{equation}
  \frac{\d \rho_{j}}{\d t} = \sum_{i} m_{i} \left( \vec{v}_{ji} \cdot
  \nabla_{i}   W_{ji}(h_{j}) + \frac{\partial W_{ji}}{\partial h_{j}} \frac{\d
    h_{j}}{\d t} \right), 
  \label{c3rhogradh}
\end{equation}
noting the extra terms compared to \eref{c3drhodt}, and where now we set
$W_{ji} = W(\vec{r}_{ji}, h_{j})$.  Noting that 
\begin{equation}
  \frac{\d h_{j}}{\d t} = \frac{\d h_{j}}{\d \rho_{j}} \frac{\d \rho_{j}}{\d
    t}, 
  \label{c3dhdt}
\end{equation}
and using \eref{c3hrho} we see that 
\begin{equation}
  \frac{\d h_{j}}{\d \rho_{j}} = - \frac{h_{j}}{3 \rho_{j}}.
\end{equation}
Substituting this into \eref{c3rhogradh} and gathering like terms, we find that
the time derivative of the density becomes 
\begin{equation}
  \frac{\d \rho_{j}}{\d t} = \frac{1}{\Omega_{j}} \sum_{i} m_{i} \vec{v}_{ji}
  \cdot \nabla_{i} W_{ij}(h_{j}) 
  \label{c3drhodtvarh}
\end{equation}
where 
\begin{eqnarray}
  \Omega_{j} 
  &=& 1 - \frac{\d h_{j}}{\d \rho_{j}} \sum_{i} m_{i} \frac{\partial
    W_{ji}(h_{j})}{\partial h_{j}}, \\
  \label{c3Omega}
  &=& 1 + \frac{h_{j}}{3 \rho_{j}} \sum_{i} m_{i} \frac{\partial
    W_{ji}(h_{j})}{\partial h_{j}},
\end{eqnarray}
and where $\partial W_{ji}/ \partial h_{j}$ is known from the choice of
kernel.  Although it can be calculated directly from the kernel, in the case
of the cubic spline kernel given in \eref{c3splinekernel} it is generally
evaluated by noting that 
\begin{equation}
  \frac{\partial W}{\partial h_{j}} = - x \nabla W - \frac{3}{h} W,
\end{equation}
where $W$ and $\nabla W$ are given by
\erefs{c3splinekernel}{c3gradsplinekernel} respectively.

Similarly, there is a correction factor to the momentum equation to allow for
the spatial variation in smoothing lengths.  Recall from \eref{c3dLdr1} that
in order to calculate the spatial variation of the Lagrangian, we need to know
the spatial derivative of the density.  Allowing now for variable smoothing
lengths and using \eref{c3drhodr}, we therefore find that 
\begin{equation}
  \frac{\partial \rho_{j}}{\partial \vec{r}_{i}} = \sum_{k} m_{k} \left(
    \nabla_{j} W_{ji}(h_{j}) \left[ \delta_{ji} - \delta_{jk} \right] +
    \frac{\partial W_{jk}(h_{j})}{\partial h_{j}} \frac{\d h_{j}}{\d \rho_{j}}
    \frac{\partial \rho_{j}}{\partial \vec{r}_{i}} \right).
\end{equation}
By gathering like terms, we find that the correction factor for the spatial
derivative of the density is same as that for the temporal one, namely that
\begin{equation}
  \frac{\partial \rho_{j}}{\partial \vec{r}_{i}} = \frac{1}{\Omega_{j}}
  \sum_{k} m_{k} \nabla_{i} W_{jk}(h_{j}) \left[ \delta_{ji} - \delta_{jk}
    \right], 
  \label{c3drhodrvarh}
\end{equation}
with the factor $\Omega_{j}$ defined as before in \eref{c3Omega}.  

Following the same derivation as in \sref{c3MomentumConservation}, it is then
easy to show that the acceleration due to hydrodynamic forces with spatially
varying smoothing lengths is given by
\begin{equation}
  \frac{\d \vec{v}_{j}}{\d t} = - \sum_{i} m_{i} \left(
  \frac{P_{j}}{\Omega_{j} \rho_{j}^{2}} \nabla_{i} W_{ji}(h_{j}) +
  \frac{P_{i}}{\Omega_{i} \rho_{i}^{2}} \nabla_{j} W_{ji}(h_{i}) \right).
  \label{c3eqnmotionvarh}
\end{equation}
Finally, from \eref{c3dUdtdef}, we see that the evolution of the internal
energy in the presence of variable smoothing lengths becomes
\begin{equation}
  \frac{\d u_{j}}{\d t} = \frac{P_{j}}{\Omega_{j} \rho_{j}^{2}}
  \sum_{i} m_{i} \vec{v}_{ji} \cdot \nabla W_{ji}(h_{j}).
  \label{c3dUdtvarh}
\end{equation}
By an analogous process to that described in \sref{c3energyconservation}, it
is possible to show that this equation for the evolution of the internal
energy is also explicitly conservative of the total energy of the system,
$E$.  The three \erefss{c3densitygradh}{c3eqnmotionvarh}{c3dUdtvarh} along with
the relationship between the density and the smoothing length \eref{c3hrho}
therefore form a fully consistent, fully conservative SPH formalism with
spatially varying smoothing lengths. 

A problem exists however, in that in order to obtain the density, one needs to
know the smoothing length (\eref{c3densitygradh}) and to obtain the smoothing
length one needs to know the density (\eref{c3hrho}).  In order to resolve
this, this pair of equations can be solved iteratively by the Newton-Raphson
method \citep{PriceMonaghan07}.  By rewriting \eref{c3hrho}, we can combine
these two equations to reduce the problem to that of finding the root $h_{j}$
of the equation $\zeta(h_{j}) = 0$, where  
\begin{equation}
  \zeta(h_{j}) = m_{j} \left( \frac{\eta}{h_{j}} \right)^{3} - \sum_{i} m_{i}
  W(\vec{r}_{ji}, h_{j}).
\end{equation}
Here the first term represents the density obtained from assuming a fixed mass
within the smoothing kernel, while the second term is the standard SPH
estimate for the density.  From some initial estimate of the root $h_{j}$, the
Newton-Raphson method gives a better estimate as being
\begin{equation}
  h_{j,\mathrm{new}} = h_{j} - \frac{\zeta(h_{j})}{\zeta'(h_{j})},
\end{equation}
where the prime denotes differentiation with respect to $h$.  By using
\eref{c3Omega} we see that 
\begin{equation}
  \zeta'(h_{j}) = -\frac{3 \rho_{j} \Omega_{j}}{h_{j}},
\end{equation}
and thus the updated value $h_{j,\mathrm{new}}$ is given by
\begin{equation}
  h_{j,\mathrm{new}} = h_{j} \left( 1 + \frac{\zeta(h_{j})}{3 \rho_{j}
    \Omega_{j}} \right).
\end{equation}
This may be repeated until $|h_{j,\mathrm{new}} - h_{j}| /h_{j} < \epsilon$
for some small value of $\epsilon$, frequently set to $10^{-3}$.  Then in
turn, a self consistent value of the density is then obtained from
\eref{c3hrho}.  As there is generally relatively little change in $h_{j}$ and
$\rho_{j}$ between timesteps, the estimator for $h_{j}$ is taken as the value
from the previous timestep, and convergence usually occurs within a small
number of iterations \citep{PriceMonaghan07}.  In pathological cases where the
Newton-Raphson method does not converge, other, universally convergent but
slower methods such as the bisection method may be used instead.

Although the inclusion of variable smoothing lengths through this method does
inevitably increase the computational cost of the code, this is relatively
small, and the conservation properties are recovered to within machine (and
integrator) tolerance.  Other tricks, such as predicting the change in the
smoothing length using \eref{c3dhdt} can reduce the computational cost still
further (see for instance, \citealt{PriceMonaghan07}).

\section{Including Gravity}
\label{c3gravity}

As many astrophysical situations are driven at some level by gravitational
forces, it is important to be able to include this consistently within the SPH
framework, and in such a manner that the inherent conservation properties of
the algorithm are not compromised.  While much work has been put into
N-body simulations of discrete particles, within the SPH formalism we
are aiming to model the gravitational force over a continuum, and thus
it should be smoothed (or \emph{softened} in SPH parlance) in a
similar manner to that in which the discrete particle mass is smoothed
into the density of a fluid continuum.  In this section we therefore consider
how this can be done in a consistent manner, and one in which as before
momentum and energy are explicitly conserved.

\subsection{Gravity in the Lagrangian}
In an extension to the Lagrangian for the hydrodynamic equations of motion, it
is possible to incorporate the effects of gravity by considering a Lagrangian
of the form
\begin{equation}
  \mathcal{L}(\vec{r}, \vec{v}) = \sum_{i} m_{i} \left(
  \frac{1}{2} \vec{v}_{i} \cdot \vec{v}_{i} -
  u_{i}(\vec{r}_{i})\right)  - \Psi,
  \label{c3Lagrangiangrav}
\end{equation}
where $\Psi$ is an as yet undefined measure of the total gravitational
potential energy of the system.  By comparison with
\eref{c3discreteLagrangian} this is clearly just the hydrodynamic Lagrangian
with an additional term  
\begin{equation}
  \mathcal{L}_{\mathrm{grav}} =  - \Psi
  \label{c3Lgrav}
\end{equation}
which describes the effects of gravity.  

Now, as with the density, at position $\vec{r}_{i}$ we can obtain the local 
gravitational potential $\Phi_{i}$, via a sum over all particles such that 
\begin{equation}
  \Phi_{i} = G \sum_{j} m_{j} \phi(\vec{r}_{i} - \vec{r}_{j},\varepsilon_{i}),
  \label{c3Phi}
\end{equation}
where $\phi(\vec{r}_{i} - \vec{r}_{j}) = \phi(\vec{r}_{ij})$ is known as the
(gravitational) softening kernel, $G$ is the universal gravitational constant
and where $\varepsilon_{i}$ is the softening length associated with particle
$i$.  The softening kernel at this stage is fairly general, but it must have
the following properties: 
\begin{itemize}
\item{$\phi(r,h) < 0$ for all $r,h$, as the local potential $\Phi$ must be
  strictly negative definite;}
\item{$\nabla \phi(0,h) = 0$, such that the gravitational force exerted by any
  particle on itself is zero;}
\item{$\displaystyle \lim_{r/h \rightarrow \infty}{\phi(r,h) =
    -\frac{1}{r}}$, i.e. the softening should reduce to zero at large
  inter-particle distances, and the Newtonian potential should be recovered.}
\end{itemize}
Generally speaking, and throughout this thesis, it is assumed that the
softening length is exactly equal to the smoothing length for all particles,
i.e. $\varepsilon_{i} = h_{i}$ for all $i$.  In a similar manner it is
generally taken that the force should only be softened when $r < 2h$, so that
force softening and density smoothing occur over exactly the same region. 

Noting that the gravitational potential energy is just the mass times the
gravitational potential, since the latter is defined over pairs of particles,
by definition the \emph{total} gravitational potential energy of the system is
given by the sum over all \emph{pairs} of particles, such that 
\begin{eqnarray}
  \Psi &=&  G \sum_{i} m_{i} \sum_{j \leq i} m_{j} \phi_{ij}(h_{i})
  \label{c3psi1} \\
  &=&  \frac{G}{2} \sum_{i} \sum_{j} m_{i} m_{j} \phi_{ij}(h_{i})
  \label{c3psi2}
\end{eqnarray}
Note that \eref{c3psi1} sums over all pairs of particles, including the
so-called self-interaction terms where $i=j$, and thus explains the factor of
a half in \eref{c3psi2}.  From this definition we therefore find that 
\begin{equation}
  \Psi = \frac{1}{2} \sum_{i} m_{i} \Phi_{i},
\end{equation}
and thus that the full Lagrangian in the presence of gravity becomes
\begin{equation}
  \mathcal{L}(\vec{r}, \vec{v}) = \sum_{i} m_{i} \left(
  \frac{1}{2} \vec{v}_{i} \cdot \vec{v}_{i} -
  u_{i}(\vec{r}_{i}) - \frac{1}{2} \Phi_{i} \right).
  \label{c3Lagrangiangrav2}
\end{equation}

By considering only the gravitational term in the Lagrangian, we can as before
use the Euler-Lagrange equations \ref{c3Leqsmotion} to obtain the acceleration
due to gravity, which becomes 
\begin{equation}
  m_{j} \frac{\d v_{j}}{\d t} = \frac{\partial \mathcal{L}_{\mathrm{grav}}}
  {\partial \vec{r}_{j}}.
  \label{c3Leqsmotiongrav}
\end{equation}
Using \erefs{c3Lgrav}{c3psi2} we therefore find that the spatial derivative of
the gravitational Lagrangian becomes 
\begin{eqnarray}
  \frac{\partial \mathcal{L}_{\mathrm{grav}}}{\partial \vec{r}_{j}} &=&
  -\frac{G}{2} \sum_{i} \sum_{k} m_{i} m_{k} \, \frac{\partial
    \phi_{ik}(h_{i})}{\partial \vec{r}_{j}}\\  
  &=& -\frac{G}{2} \sum_{i} \sum_{k} m_{i} m_{k} \left( \nabla_{j}
  \phi_{ik}(h_{i}) + \frac{\partial \phi_{ik}(h_{i})}{\partial h_{i}}
  \frac{\partial h_{i}}{\partial \vec{r}_{j}}  \right).
  \label{c3dLgravdr}
\end{eqnarray}
Here we see that in the case of fixed smoothing (and therefore softening)
lengths, $\partial h_{i}/ \partial \vec{r}_{k} = 0$, and thus we only require
the first term to determine the effects of gravity. The second term is
therefore a correction term to allow for spatial variation in $h$.

As before, using the method of equations \ref{c3drhodr1} to \ref{c3drhodr}
the spatial gradient becomes
\begin{eqnarray}
  \left. \frac{\partial \mathcal{L}_{\mathrm{grav}}}{\partial \vec{r}_{j}}
  \right|_{h} 
  &=& -\frac{G}{2} \sum_{i} \sum_{k} m_{i} m_{k} \, \nabla_{j}
  \phi_{ik}(h_{i}) \left[\delta_{ij} - \delta_{kj} \right],  \\   
  &=& -\frac{G}{2} m_{j} \, \sum_{k} m_{k} \nabla_{j} \phi_{jk}(h_{j})
  + \frac{G}{2} \sum_{i} m_{i} m_{j} \nabla_{j} \phi_{ij}(h_{i}).
\end{eqnarray}
Now by changing the summation index of the first term to $i$, and noting again
that the kernel is antisymmetric we obtain
\begin{equation}
  \left. \frac{\partial \mathcal{L}_{\mathrm{grav}}}{\partial \vec{r}_{j}}
  \right|_{h} = -\frac{G}{2} m_{j} \sum_{i} m_{i} \left( \nabla_{j}
  \phi_{ji}(h_{j}) + \nabla_{j} \phi_{ji}(h_{i}) \right),
\end{equation} 
which therefore encapsulates the effects of gravity in the case of constant
smoothing lengths.

If we now consider the second term in \eref{c3dLgravdr} and self-consistently
correct for spatial variation in the smoothing length, we find that 
\begin{equation}
  \left. \frac{\partial \mathcal{L}_{\mathrm{grav}}}{\partial \vec{r}_{j}}
  \right|_{\mathrm{corr}} = 
  -\frac{G}{2} \sum_{i} \sum_{k} m_{i} m_{k} \, \frac{\partial
    \phi_{ik}(h_{i})}{\partial h_{i}} \frac{\d h_{i}}{\d \rho_{i}}
  \frac{\partial \rho_{i}}{\partial \vec{r}_{j}}.
\end{equation}
By substituting \eref{c3drhodrvarh} into the above we find that 
\begin{align}
    \left. \frac{\partial \mathcal{L}_{\mathrm{grav}}}{\partial \vec{r}_{j}}
  \right|_{\mathrm{corr}} 
  =&  -\frac{G}{2} \sum_{i} \sum_{k} m_{i} m_{k} \, \frac{\partial
    \phi_{ik}(h_{i})}{\partial h_{i}} \frac{\d h_{i}}{\d \rho_{i}}
  \frac{1}{\Omega_{i}} \sum_{l} m_{l} \nabla_{j} W_{il}(h_{l}) [\delta_{ij} -
    \delta_{lj}] \\
  \begin{split}
  =& -\frac{G}{2} m_{j} \sum_{k} m_{k} \frac{\partial \phi_{jk}}{\partial
    h_{j}} \frac{\d h_{j}}{\d \rho_{j}} \frac{1}{\Omega_{j}} \sum_{l} m_{l}
  \nabla_{j} W_{jl}(h_{l}) \\
  & \quad  + \frac{G}{2} \sum_{i} \sum_{k} \frac{\partial
    \phi_{ik}(h_{i})}{\partial h_{i}} \frac{1}{\Omega_{i}} m_{j} \nabla_{j}
  W_{ij}(h_{i}).
  \end{split}
  \label{c3gravvarh1}
\end{align}
Now by changing the summation index of the second sum in the first term of
\eref{c3gravvarh1} to $i$, defining a new quantity $\xi_{p}$ such that 
\begin{equation}
  \xi_{p} = \frac{\d h_{p}}{\d \rho_{p}} \sum_{q} m_{q} \frac{\partial
    \phi_{pq}(h_{p})}{\partial h_{p}},
  \label{c3xi}
\end{equation}
and using the antisymmetry property of the gradient of the smoothing kernel,
we see that the correction term reduces to
\begin{equation}
  \left. \frac{\partial \mathcal{L}_{\mathrm{grav}}}{\partial \vec{r}_{j}}
  \right|_{\mathrm{corr}} =  -\frac{G}{2} m_{j} \sum_{i} m_{i} \left(
  \frac{\xi_{j}}{\Omega_{j}} \nabla_{j} W_{ji}(h_{j}) +
  \frac{\xi_{i}}{\Omega_{i}} \nabla_{j}W_{ji}(h_{i}) \right).
  \label{c3gravvarh}
\end{equation}

Finally, using \eref{c3Leqsmotiongrav} and by incorporating the effects of
gravity into the equations of motion for a hydrodynamic flow with artificial
viscosity (while self-consistently allowing for variable smoothing lengths) we
find that the full equations of motion become 
\begin{equation*}
  \frac{\d \vec{v}_{j}}{\d t} 
  = - \sum_{i} m_{i} \left( \frac{P_{j}}{\Omega_{j} \rho_{j}^{2}} \nabla_{j}
  W_{ji}(h_{j}) + \frac{P_{i}}{\Omega_{i} \rho_{i}^{2}} \nabla_{j}
  W_{ji}(h_{i}) + \Pi_{ji} \frac{\nabla_{j}W_{ji}(h_{j}) +
    \nabla_{j}W_{ji}(h_{i})}{2} \right)
\end{equation*}
\vspace{-5mm}
\begin{equation}
  \quad - \frac{G}{2} \sum_{i} m_{i} \, \left( \nabla_{j} \phi_{ji}(h_{j}) +
  \nabla_{j} \phi_{ji}(h_{j}) \right) \qquad \qquad \qquad \qquad \quad
  \label{c3eqnmotionvarhgrav}
\end{equation}
\vspace{-3mm}
\begin{equation*}
  - \frac{G}{2} \sum_{i} m_{i} \left(
  \frac{\xi_{j}}{\Omega_{j}} \nabla_{j} W_{ji}(h_{j}) +
  \frac{\xi_{i}}{\Omega_{i}} \nabla_{j}W_{ji}(h_{i}) \right), \qquad \qquad
\end{equation*}
with $\Omega_{i}$ and $\xi_{i}$ defined as per \erefs{c3Omega}{c3xi}
respectively\footnote{Note that for consistency, the artificial viscosity term
  $\Pi_{ji}$ uses the \emph{average} value of the smoothing lengths $h_{ji}
  = (h_{j} + h_{i})/2$ in its definition of $\mu_{ji}$ (\eref{c3mu}).}.  As in
the case for the pure hydrodynamic flow, the use of a Lagrangian in deriving
these equations ensures the explicit conservation of both linear and angular
momentum, which is also clear from the pairwise symmetry present in all terms
in the above equation.

\subsection{Evolution of the Gravitational Potential}
Clearly as particles move about within a gravitational potential, their
potential energy (given in SPH terms by $m_{j} \Phi_{j}$) will also vary.
Although the potential (and thus the potential energy) is obtained at any
point by the sum over particles using \eref{c3Phi}, the time evolution of the
potential energy is required to maintain energy conservation.  Hence in a
similar manner to \sref{c3energyconservation} we must consider the total
energy of the system, which including the gravitational potential energy
becomes 
\begin{equation}
  E = \sum_{j} m_{j} \left( \frac{1}{2} \vec{v}_{j} \cdot \vec{v}_{j} +
  u_{j} + \frac{1}{2} \Phi_{j} \right).
  \label{c3gravenergybalance}
\end{equation}
As before, to ensure energy conservation we require that the time derivative
of the total energy is zero, i.e. that
\begin{equation}
  \sum_{j} m_{j} \left(  \vec{v}_{j} \cdot \frac{\d \vec{v}_{j}}{\d t} +
  \frac{\d u_{j}}{\d t} + \frac{1}{2} \frac{\d \Phi_{j}}{\d t}
  \right) = 0. 
\end{equation}
By considering \eref{c3Phi}, we see that 
\begin{equation}
  \frac{\d \Phi_{j}}{\d t} = \frac{G}{2} \sum_{i} m_{i} \left( \nabla_{j}
    \phi_{ji}(h_{j}) \cdot \frac{\d \vec{r}_{j}}{\d t} + \nabla_{i}
    \phi_{ji}(h_{j}) \cdot \frac{\d \vec{r}_{i}}{\d t} + \frac{\partial
      \phi_{ji}}{\partial h_{j}} \frac{\d h_{j}}{\d \rho_{j}} \frac{\d
      \rho_{j}}{\d t} \right).
\end{equation}
Recalling the definition of $\xi_{j}$ from \eref{c3xi}, and using
\eref{c3drhodtvarh} for the definition of $\d \rho_{j} /\d t$ with variable
smoothing lengths, we obtain
\begin{equation}
  \frac{\d \Phi_{j}}{\d t} = \frac{G}{2} \sum_{i} m_{i} \vec{v}_{ji} \cdot
  \left(\nabla_{j} \phi_{ji}(h_{j}) + \frac{\xi_{j}}{\Omega_{j}} \nabla_{j}
  W_{ji}(h_{j}) \right),
  \label{c3dPhidt}
\end{equation}
From \srefs{c3energyconservation}{c3smoothinglengths} we know that in the
energy balance (\eref{c3gravenergybalance}), over all particles the
hydrodynamic terms in the equations of motion (\eref{c3eqnmotionvarh}) exactly
counteract the temporal rate of change of the internal energy, 
\begin{equation}
  \frac{d E_{\mathrm{hydro}}}{\d t} = \sum_{j} m_{j} \left( \left. \vec{v_{j}}
  \cdot \frac{\d \vec{v}_{j}}{\d t} \right|_{\mathrm{hydro}} + \frac{\d
      u_{j}}{\d t} \right) = 0,
\end{equation}
and thus pure hydrodynamic flows are exactly conservative of energy.  With the
inclusion of gravity we therefore only need to show that over the whole system
the gravitational terms in the equations of motion balance the time derivative
of the gravitational potential, i.e. that 
\begin{equation}
  \frac{d E_{\mathrm{grav}}}{\d t} = \sum_{j} m_{j} \left( \left. \vec{v_{j}}
  \cdot \frac{\d \vec{v}_{j}}{\d t} \right|_{\mathrm{grav}} + \frac{1}{2}
  \frac{\d \Phi_{j}}{\d t} \right) = 0 
\end{equation}
in order to maintain exact conservation of energy in self-gravitating systems.

From \erefs{c3eqnmotionvarhgrav}{c3dPhidt} this gravitational energy balance
becomes  
\begin{equation}
  \begin{split}
    \frac{d E_{\mathrm{grav}}}{\d t} \; =
    & \; -\frac{G}{2} \sum_{j} \sum_{i} m_{j} m_{i} \vec{v}_{j} \cdot \biggl(
    \nabla_{j} \phi_{ji}(h_{j}) + \nabla_{j} \phi_{ji}(h_{i}) + \biggr. \\
    & \qquad \qquad \qquad \qquad \qquad 
    \left. \frac{\xi_{j}}{\Omega_{j}} \nabla_{j} W_{ji}(h_{j}) +
    \frac{\xi_{i}}{\Omega_{i}} \nabla_{j} W_{ji}(h_{i}) \right) \\    
    & \quad + \frac{G}{2} \sum_{j} \sum_{i} m_{j} m_{i} \, \vec{v}_{ji} \cdot
    \left( \nabla_{j} \phi_{ji}(h_{i}) +     \frac{\xi_{j}}{\Omega_{j}}
    \nabla_{j} W_{ji}(h_{j}) \right) 
  \end{split}
\end{equation}
Cancelling like terms, this reduces to 
\begin{equation}
  \begin{split}
    \frac{d E_{\mathrm{grav}}}{\d t} \; =
    & \; -\frac{G}{2} \sum_{j} \sum_{i} m_{j} m_{i} \biggl( \vec{v}_{j} \cdot
    \nabla_{j} \phi_{ji}(h_{j}) + \vec{v}_{i} \cdot \nabla_{j} \phi_{ji}(h_{i})
    \biggr. \\
    & \qquad \qquad \qquad + \left. \frac{\xi_{i}}{\Omega_{i}} \vec{v}_{j} \cdot
      \nabla_{j} W_{ji}(h_{i}) + \frac{\xi_{j}}{\Omega_{j}} \nabla_{j}
      W_{ji}(h_{j}) \right),
  \end{split}
\end{equation}
and finally, noticing that the gradients of both the smoothing and the
softening kernels are antisymmetric under a reversal of the summation indices
$i$ and $j$, we obtain the desired result that
\begin{equation}
  \frac{d E_{\mathrm{grav}}}{\d t} = 0.
\end{equation}
Therefore, we see that gravity can be included into SPH in such a manner that
the algorithm remains explicitly conservative of energy.

\subsection{Gravitational Potentials and the Softening Kernel}
Finally for this section, we need to consider the form of the gravitational
softening kernel $\phi$, and its relation to the smoothing kernel $W$.  Recall
that Poisson's equation links the gravitational potential $\Phi(\vec{r})$ to
the density $\rho(\vec{r})$ at position $\vec{r}$, such that 
\begin{equation}
  \nabla^{2} \Phi(\vec{r}) = 4 \pi G \rho(\vec{r}).
  \label{c3Poisson}
\end{equation}
Given that we implicitly assume each particle to be spherically symmetric, by
using spherical polar co-ordinates and substituting \erefs{c3density}{c3Phi}
into \eref{c3Poisson} we find that (for a generalised radial co-ordinate $r$)
\begin{equation}
  W(r,h) = \frac{1}{4 \pi r^{2}} \frac{\partial}{\partial r} \left( r^{2}
  \frac{\partial \phi(r,h)}{\partial r} \right),
\end{equation}
where we have neglected the spatial variation of $h$\footnote{This is because
  the smoothing length essentially acts as a normalising constant in both the
  smoothing and the softening kernels, and for any given particle is held
  constant within Poisson's equation.  Thus its spatial variation is
  immaterial here.}.

We can now integrate this, to link the derivative of the softening kernel
$\partial \phi / \partial r$ (also known as the force kernel) to the smoothing
kernel, such that  
\begin{equation}
  \frac{\partial \phi}{\partial r} = \frac{4\pi}{r^{2}} \int^{r} r'^{2}
  W(r') \d r' + \frac{C_{1}}{r^{2}},
\end{equation}
with the integration constant $C_{1}$ subject to the condition that for $r \geq
2h$ we recover the standard Newtonian inverse square law, which using our
definitions becomes $\partial \phi /\partial r = 1/r^{2}$.  In a similar manner
we can integrate this a step further (by parts), to give the full softening
kernel, such that    
\begin{equation}
  \phi = 4\pi \left[ -\frac{1}{r} \int^{r} r'^{2} W(r') \d r' +
    \int^{r} r' W(r') \d r' \right] + \frac{C_{1}}{r^{2}} +
  \frac{C_{2}}{r}, 
\end{equation}
where the second integration constant allows the correct asymptotic behaviour
(i.e. $\phi \rightarrow 0$ as $r \rightarrow \infty$) to be established.

With this in mind, for the cubic spline kernel defined in
\eref{c3splinekernel} the force kernel $\partial \phi /\partial r$ becomes
\begin{equation}
  \frac{\partial \phi(r,h)}{\partial r} = \left\{
  \begin{array}{lr}
    \displaystyle \frac{1}{h^{2}} \left( \frac{4}{3}x - \frac{6}{5}x^{3} +
    \frac{1}{2} x^{4} \right) & 0 \leq x \leq 1,\\
    \displaystyle \frac{1}{h^{2}} \left( \frac{8}{3}x - 3x^{2} +
    \frac{6}{5}x^{3} - \frac{1}{6}x^{4} - \frac{1}{15x^{2}} \right) & 1 \leq x
    \leq 2, \\   
    \displaystyle \frac{1}{r^{2}} & x \geq 2,
  \end{array} 
  \right.
  \label{c3forcekernel}
\end{equation}
where $x = r/h$ and where the integration constants have been absorbed to
ensure piecewise continuity.  Finally, we therefore find the full softening
kernel consistent with the cubic spline smoothing kernel to be 
\begin{equation}
  \phi(r,h) = \left\{
  \begin{array}{lr}
    \displaystyle \frac{1}{h} \left( \frac{2}{3}x^{2} - \frac{3}{10}x^{3} +
    \frac{1}{10}x^{5} - \frac{7}{5} \right) & 0 \leq x \leq 1, \\
    \displaystyle \frac{1}{h} \left( \frac{4}{3}x^{2} - x^{3} +
    \frac{3}{10}x^{4} - \frac{1}{30} x^{5} - \frac{8}{5} + \frac{1}{15x}
    \right) & 1 \leq x \leq 2, \\
    \displaystyle -\frac{1}{r} & x \geq 2.
    \end{array}
  \right.
  \label{c3gravsoftkernel}
\end{equation}
Using this definition of the softening kernel along with the cubic spline
smoothing kernel \eref{c3splinekernel}, the equations of motion
\eref{c3eqnmotionvarhgrav} and \eref{c3dPhidt} for the evolution of the
gravitational potential therefore allows gravity to be included in a manner
that it is fully conservative, and is such that Poisson's equation is
satisfied throughout.
 
\section{Finding the Nearest Neighbours}
\label{c3neighbours}

Various methods exist for finding the nearest neighbours (i.e.  those particles
within the smoothing kernel of any given particle), with the simplest being a
direct search over all particles.  This becomes very expensive in the limit of
large numbers of particles $N$ however, as the computational cost scales as
$\ord{N^2}$.  Other methods such as using an overlaid grid or a linked list of
particle positions have been used
\citep{HockneyEastwood81,Monaghan85,Murray96,DeeganPhD}.  One of the more
efficient methods however is to use a hierarchical tree structure, an approach
that grew out the requirements of N-body codes to distinguish distant
particles (where the gravitational forces could be evaluated via multipole
expansions) from local particles (where direct N-body calculation of the
forces was still required).  These in general reduce the cost of
neighbour-finding from $\ord{N^{2}}$ to $\ord{N \log N}$
\citep{BarnesHut86,Hernquist87,HernquistKatz89}, although reductions to
$\ord{N}$ have been achieved \citep{Dehnen02,Dehnen00}. 

Trees are essentially data structures which decompose the computational domain
into a series of discrete volumes, the sum over which contains all the
particles.  The smallest volumes generally contain only a single particle, but
this is not strictly necessary, and indeed may inhibit the efficiency of the
code \citep[private communication]{DehnenPC}.  By construction, particles
which are near each other in space are near each other within the tree
structure, and thus by looping over a relatively small part of the tree, the
nearest neighbours may be found efficiently.  There are various different
algorithms that perform this decomposition described in the literature, none
of which are trivial, so I shall not attempt to go into any depth here.  For
further details, see for instance
\citet{NumericalRecipes3,Dehnen02,SteinmetzMuller93,BarnesHut86,Bentley75} and
references therein.   

A distinct advantage of using trees for neighbour finding within an SPH code
is that they couple readily with pre-existing methods for evaluating the
gravitational force between large numbers of particles.  Rather than a direct
summation (which is of $\ord{N^{2}}$) over all particles to find the
gravitational force at a specific location, particles at large distances can
be effectively treated as a single body, and multipole expansions used to
approximate the force.  This approach has found success in various \Nbody
codes as a means of reducing the computation time to $\ord{N \log N}$ or lower
\citep{BarnesHut86,Hernquist87,HernquistKatz89}.  Use of a tree algorithm
therefore allows the process of neighbour finding to be coupled to that of
finding the gravitational forces acting on a particle, with an attendant
saving of computational expense.

\section{Integration and Timestepping}
\label{c3timesteps}

So far we have obtained equations to evolve the density, the three components
of velocity under the influence of pressure, gravitational and (artificial)
viscous forces, the internal energy and the gravitational potential.  Finally
therefore it is time to consider \emph{how} these equations are actually
evolved, and to discuss the issues of temporal integration and time-stepping.

Generally speaking there are two principal methods used to perform the time
evolution, and indeed the code I have used throughout includes the option to
use either.  They are the so-called Leapfrog integrator (also known as the
kick-drift or St\"ormer-Verlet integrator) and the Runge-Kutta-Fehlberg method,
and I shall now briefly consider both of these. 

\subsection{The Leapfrog Integrator}
The leapfrog integrator is a second-order integrator, so-called because the
position and the velocity are advanced half a timestep out of phase, with each
update of position or velocity using the value of the velocity or position
evaluated at the previous half timestep.  In this manner, the positions  and
velocities ``leap-frog'' over each other at every half timestep, giving rise
to the name.  The leapfrog method is widely used in \Nbody codes, since in the
case where the acceleration is independent of the velocity, i.e. where
$\vec{a} = \vec{a}(\vec{r})$ only, it is particularly simple to implement.  In
its ``pure'' form it is a time-reversible, symplectic integrator, which by
definition is explicitly conservative of both energy and angular momentum (see
for instance \citet{Springel05} and references therein).

In essence then, if the position, velocity and acceleration at time $t_{i}$
are given by $\vec{r}_{i}$, $\vec{v}_{i}$ and $\vec{a}_{i}$ respectively, with
a timestep $\delta t$ the standard form of the leapfrog integrator gives the
positions and velocities as 
\begin{equation}
  \begin{split}
    \vec{r}_{i+1}   &= \vec{r}_{i} + \vec{v}_{i-1/2} \, \delta t, \\
    \vec{v}_{i+1/2} &= \vec{v}_{i-1/2} + \vec{a}_{i} \, \delta t.
    \label{c3LFhalfdt}
  \end{split}
\end{equation}
Here it is clear that the positions and velocities are evaluated at half
timesteps with respect to each other, and ``leap-frog'' over each other as
they are evolved.  In this form it is also clear that the integrator should be
perfectly time-reversible. 

A form that is often more readily applied is the equivalent definition at
integer timesteps, which becomes 
\begin{equation}
  \begin{split}
     \vec{r}_{i+1} & = \vec{r}_i + \delta t \left( \vec{v}_{i} +
     \frac{\delta t}{2} \, \vec{a}_{i} \right),\\
     \vec{v}_{i+1} &= \vec{v}_i + \frac{\delta t}{2} \left(\vec{a}_i +
       \vec{a}_{i+1}\right).
     \label{c3LFintegerdt}
   \end{split}
 \end{equation}
Although it is now less obvious, these equations are still fully time
reversible.  Note further that the form of the increments on the RHS of each
of the above equations is equivalent to an estimate of the relevant quantity
at the following half timestep, noting in particular that 
\begin{equation}
 \vec{v}_{i} + \frac{\delta t}{2} \, \vec{a}_{i} = \vec{v}_{i+1/2},
  \label{c3LFvpredictor}
\end{equation}
to first order.

Note however, that in both cases problems arise if the acceleration depends on
the velocity, since from \eref{c3LFintegerdt} we see that to calculate
$\vec{v}_{i+1}$ we already need to know the acceleration $\vec{a}_{i+1}$, and
the scheme becomes implicit.  Since in SPH simulations both the pressure force
and the artificial viscous force depend on the local velocity, it is clear
that modifications are required before this integrator may be used.  The
standard way this correction is implemented (see for instance 
\citealt{Springeletal01,VINEpaper1}) is as follows: 
\begin{itemize}
  \item[-]{Firstly, predict the positions at time $t_{i+1/2}$ in a
      manner analogous to \eref{c3LFvpredictor} via 
      \begin{equation}
        \vec{r}_{i+1/2} = \vec{r}_{i} + \frac{\delta t}{2} \, \vec{v}_{i}.
      \end{equation}
      }
    \item[-]{Secondly, use \eref{c3LFvpredictor} to obtain the velocity
      at time $t_{i+1/2}$, and extrapolate other values (such as
      density, internal energy and gravitational potential) at the
      half timestep also.  Hence calculate the acceleration at the
      half timestep, $\vec{a}_{i+1/2}$.}     
    \item[-]{Calculate the velocity at time $t_{i+1}$ using  
        \begin{equation}
          \vec{v}_{i+1} = \vec{v}_{i} + \delta t \, \vec{a}_{i+1/2}.
        \end{equation}
      }
    \item[-]{Now update the positions to timestep $t_{i+1}$ using
      \begin{equation}
        \vec{r}_{i+1} = \vec{r}_{i} + \frac{\delta t}{2} \left(
        \vec{v}_{i} + \vec{v}_{i+1} \right).
      \end{equation}
    }
\end{itemize}
The process is now repeated as required.  

Although the strict symmetry between the integration of positions and
velocities has been lost by the inclusion of these predictor steps, this
method still remains time-reversible.  Furthermore, it is also now possible to
include adaptive timestepping, which would have been problematic before
precisely \emph{because} of the symmetry between the integrations
\citep{VINEpaper1}.  Generally speaking however, maintaining
time-reversibility with adaptive timestepping is difficult
\citep{Springeletal01,Quinnetal97}, though not impossible. 

\subsection{The Runge-Kutta-Fehlberg Integrator}
Runge-Kutta methods for integrating systems of differential equations are well
known, tried and trusted methods, which use multiple estimates of the
derivative across a given timestep to arrive at accurate, generally high order
estimates for the evolved quantities.  Most common is the fourth order
Runge-Kutta method, often simply abbreviated to RK4, which has been known and
used for over a century \citep{Kutta1901}.  Moreover,
\citet{Fehlberg68,Fehlberg69} obtained a modified Runge-Kutta integrator (now
known as a Runge-Kutta-Fehlberg integrator) which embedded a order $n+1$
method within an order $n$ method.  This allows the two methods to be compared
to give an error estimate, and thus for the error to be controlled to some
given tolerance.  The most common of these methods embeds a fifth order estimate
within a fourth order scheme, and is therefore known as an RK45 integrator. 

However, compared to the leapfrog integrator, which requires only one
evaluation of the acceleration per timestep, the RK4 scheme requires four, and
the RK45 method requires six.  Therefore these methods, although correct to
much higher order than the leapfrog, add significantly to the computation time
required.  (Note also that they are not necessarily more \emph{accurate}
either, as they are not explicitly conservative in the way that the leapfrog
method is.  See for instance \citet{Springel05,VINEpaper1,Rosswog09} for a
comparison of these integrators as applied to a simple Keplerian orbit evolved
over many dynamical times.)  The solution is to go to a lower order RKF
method, where the implicit error control is still present but the number of
derivative evaluations is reduced.  A common choice for many SPH codes
including VINE \citep{VINEpaper1,VINEpaper2} and the one used in the code I
have used, is the RK12 integrator developed by \citet{Fehlberg69} which
proceeds as follows. 

For a given variable $\vec{z}$, the evolution from $\vec{z}_{i}$ at time
$t_{i}$ to $\vec{z}_{i+1}$ at time $t_{i+1} = t_{i} + \delta t$ is given by
\begin{equation}
  \vec{z}_{i+1} = \vec{z}_{i} + \left(\frac{1}{256}k_{0} +
  \frac{255}{256}k_{1} \right) \delta t,
  \label{c3RK1}
\end{equation}
where the values of $k_{0}$ and $k_{1}$ are provided by evaluating
$\dot{\vec{z}}$ at various points, such that 
\begin{equation}
  \begin{split}
    k_{0} &= \dot{\vec{z}}(t_{i}, \, \vec{z}_{i}), \\
    k_{1} &= \dot{\vec{z}}(t_{i} + \frac{\delta t}{2}, \, \vec{z}_{i} +
    \frac{\delta t}{2} k_{0} ) 
  \end{split}
\end{equation}
and where the dot as usual denotes differentiation with respect to time.
Expansion via Taylor series shows that this is accurate to first order, with
the choice of coefficients in \eref{c3RK1} producing a leading order
truncation error $\tau_{\mathrm{trunc}}$ such that  
\begin{equation}
  \tau_{\mathrm{trunc}} = -\frac{1}{512} \delta t^{2} \,
  \ddot{\vec{z}}.
  \label{c3RK1err}
\end{equation}

Using the values for $k_{0}$ and $k_{1}$ defined above, we can compute a
further estimate $\vec{z}^{*}_{i+1}$ for $\vec{z}$ at time $t_{i+1}$, such
that  
\begin{equation}
  \vec{z}^{*}_{i+1} = \vec{z}_{i} + \frac{\delta t}{2} \left( \frac{1}{512}
  k_{0} + \frac{255}{256} k_{1} + \frac{1}{512} k_{2} \right),
  \label{c3RK2}
\end{equation}
with the additional value $k_{2}$ defined such that 
\begin{equation}
  \begin{split}
    k_{2} &= \dot{\vec{z}} (t_{i} + \delta t, \, \vec{z}_{i} + \left(
    \frac{1}{256}k_{0} +  \frac{255}{256}k_{1} \right)  \delta t ), \\
    &= \dot{\vec{z}}(t_{i+1}, \vec{z}_{i+1}).
  \end{split}
\end{equation}
Again, by considering Taylor series expansions, this value $\vec{z}^{*}_{i+1}$
can be shown to be a second order estimate.  One of the more appealing tricks
of this method is that here $k_{2}$ is simply $k_{0}$ evaluated for the
\emph{next} timestep, and thus per timestep, only two derivative evaluations
are required.

We now therefore have both a first and a second order estimate for the value
of $\vec{z}$ at time $t_{i+1}$, with a known truncation error for the first
order method.  This can therefore be used for error control, to
ensure that the timestep used is appropriate (see for instance,
\citealt{NumericalRecipes3}).  However, in order for this error control to be
valid, the \emph{first} order scheme must be used for the evolution.  To
mitigate this, by construction this first order scheme has very small second
order errors (\eref{c3RK1err}), and so is effectively a quasi-second order
integrator. 

\subsection{Timestepping Criteria}
For either integrator, it is crucial that the timestep size is chosen
correctly, both to ensure the accuracy of the evolution and to ensure numerical
stability.  In this section I shall briefly discuss the principal timestepping
criteria in general use, and one specific to the code I have used.  

\subsubsection{CFL Criterion}
By far the most general timestep criterion for gas-dynamical systems is the
so-called \emph{Courant-Friedrichs-Lewy} or CFL condition, given in it
simplest form by 
\begin{equation}
  \delta t_{\mathrm{CFL}} \leq  \frac{\delta x}{c},
\end{equation}
where $\delta x$ is a characteristic length scale, and $c$ is a characteristic
speed \citep{Anderson95}.  For SPH simulations, these are both well defined;
the smoothing length $h$ provides the characteristic length, and sound speed
$\cs$ gives the characteristic speed of the medium. The CFL condition for
particle $i$ then becomes \begin{equation}
  \delta t_{\mathrm{CFL}} \leq \frac{h}{\cs}.
\end{equation}
This has a ready physical interpretation in that it prevents spatial
information transfer through the code at a rate greater than the local sound
speed.  In the presence of artificial viscosity this requires a slight
alteration, and as such \citet{BateBP95} recommend using the following
criterion;
\begin{equation}
  \delta t_{\mathrm{CFL}} = \frac{0.3h}{\cs + h|\nabla \cdot \vec{v}| + 1.2
    (\alpha_{\mathrm{SPH}} \cs + \beta_{\mathrm{SPH}} h | \nabla \cdot
    \vec{v}|)},
  \label{c3dtCFL}
\end{equation}
where the factors of 0.3 in the numerator and 1.2 in the denominator are
empirically determined.  The $\alpha_{\mathrm{SPH}}$ and
$\beta_{\mathrm{SPH}}$ terms are those used to determine the strength of the
artificial viscosity (see \sref{c3dissipation}), and it should be noted that
the final term in the denominator is only included in the case where $|\nabla
\cdot \vec{v}| < 0$.  The extra $h|\nabla \cdot \vec{v}|$ term in the
denominator accounts for the expansion or contraction of the flow, and thus
explicitly allows for compressibility effects.  There are variations on this
theme (see for instance \citealt{DeeganPhD,Monaghan92,Monaghan89}) but the
definition given above is the one present in the code I have used.

\subsubsection{Force Condition}
A further commonly used timestep condition is that based on the
acceleration of the particle, known as the force condition.  This is simple in
form, and is given by
\begin{equation}
  \delta t_{\mathrm{F}} = f_{\mathrm{F}} \sqrt{\frac{h}{|\vec{a}|}},
  \label{c3dtF}
\end{equation}
where as before $\vec{a}$ is the particle acceleration, and $f_{\mathrm{F}} < 1$
is a tuning constant.  Values for $f_{\mathrm{F}}$ vary from code to code but
are generally in the range 0.25 - 0.5 \citep{VINEpaper1,BateBP95,Monaghan89}.
The code I have used employs $f_{\mathrm{F}} = 0.3$.

\subsubsection{Integrator Limits}
Dependent on the choice of integrator, other timestep criteria may be
required.  In particular, if using the RKF method the timestep criterion
associated with the error correction must be incorporated.  Using the method
outlined above, this corresponds to a timestep of 
\begin{equation}
  \delta t_{\mathrm{RKF}} = \delta t_{\mathrm{old}} \sqrt{\frac{512
      \epsilon}{|\vec{z}_{RK2} - \vec{z}_{RK1}|}},
  \label{c3dtRKF}
\end{equation}
where $\epsilon$ is the desired error tolerance (usually of the order of
$10^{-4}$ - $10^{-5}$) and the $\vec{z}_{RK1}, \vec{z}_{RK2}$ are the
predictions for any quantity $\vec{z}$ from the first and second order methods
within the integrator respectively.  The $\delta t_{\mathrm{old}}$ term is
simply the increment used for the previous step.

\subsubsection{Generalised Timestep Criteria}
A general class of additional timestep criteria may be obtained by
dimensional analysis, in that for any time-varying quantity $z$ we may define
a characteristic timescale on which it varies as 
\begin{equation}
  \tau_{z} = \frac{z}{\dot{z}},
\end{equation}
where as usual $\dot{z}$ is the time derivative of $z$.  To ensure that this
timescale is properly resolved, we can therefore define a timestep condition
such that 
\begin{equation}
  \delta t_{z} = f_{z} \frac{z}{\dot{z}},
  \label{c3dtz}
\end{equation}
where $f_{z} < 1$ is a tuning factor.  Although seldom required in general,
a timestep criterion of this form was implemented into the code when looking
at the effects of strongly varying cooling times in Chapter 5, and will
be discussed in more detail there.

\subsection{Setting the Timestep}
There are therefore a variety of possible timestep choices, and thus to ensure
that they are all satisfied, the timestep for each particle used is the
minimum of all possibilities, i.e.
\begin{equation}
  \delta t_{i} = \min (\delta t_{\mathrm{CFL},i}, \delta t_{\mathrm{F},i},
  \delta t_{\mathrm{RKF},i}, \delta t_{z}).
  \label{c3dt}
\end{equation}
Where there are only relatively small changes in the characteristic
timescale, a standard choice is to use a global timestep $\delta
t_{\mathrm{glob}}$, which is set by the minimum of the timesteps $\delta
t_{i}$ for the individual particles, such that    
\begin{equation}
  \delta t_{\mathrm{glob}} = \min_{i} \, (\delta t_{i}).
\end{equation}
This has the advantage that all particles are evolved in lockstep, and thus
there is no `information lag' due to particles being on separate timesteps.

On the other hand, individual particle timestepping has the advantage of being
much faster and thus more computationally efficient wherever there are large
ranges in the timescales of the problem being investigated.  It can
however introduce instabilities into the integrator \citep{VINEpaper1}, and
can also lead to the phenomenon of low density particles on long timesteps
drifting into regions of high density evolving on much shorted timesteps,
leading to spurious entropy generation \citep[private
  communication]{PearcePC}. This latter effect is particularly noticeable in
tests of Sedov blasts (see for instance \citealt{Taskeretal08}), in which
small entropy-driven bubbles lead to granularity in the post-shock region.
This is a relatively uncommon phenomenon however, and occurs principally in
the case of strongly shocked systems.    

Integrator stability may be maximised (particularly for the leapfrog scheme)
by using timesteps that are integer multiples of each other, and generally
speaking, for a maximum timestep $T$, sub-timesteps will be given by
$2^{-n}T$.  The particle timesteps are therefore rounded \emph{down} to the
nearest relevant power of two in this case.  This is the case in the code I
have used for all the simulations presented in this thesis, which uses
individual particle timesteps and is only weakly shocked throughout. 
\section{Summary}
\label{c3summary}

In this chapter I have derived the SPH algorithm from first principles, and
then built it up in a series of steps to solve for pure hydrodynamical
isentropic flows, dissipational flows, and finally dissipational flows under
the influence of gravitational forces.  Additionally I have shown that it is
possible to self-consistently allow for spatially variable smoothing lengths,
which allows the algorithm to be highly adaptive with the local fluid
density, but to maintain exact conservation of mass, linear and angular
momentum and energy, to within the integrator tolerance.  In the case of
isentropic flows, entropy is also conserved by construction.  Furthermore I
have also briefly detailed various methods of finding the nearest neighbours,
and two means of evolving the fluid flow forward in time.

Since the problems I shall be considering in later chapters require only that
dissipational flow in the presence of gravity to be modelled, this is all that
I have covered here.  However this is by no means the limit for the SPH
formalism.  Much effort has been put into including additional physics such as
radiative transfer
(\citet{Nayakshinetal09,PetkovaSpringel09,Forganetal09,Bisbasetal09,Gritschnederetal09,PawlikSchaye08}
and \citet{Altayetal08} to name but few of the recent efforts) and magnetic
fields/MHD (see for
instance \citet{Price10,DolagStayszyn09,RosswogPrice07,PriceMonaghan05,PriceMonaghan04b}
and \citealt{PriceMonaghan04b}), and this will no doubt continue as computing
power steadily increases. 

As with any numerical scheme however, SPH remains an approximation to reality,
and as such reality checks are required in the form of standard tests.  These
act as calibration routines, to ensure the the results of any simulations are
physically realistic, and can be relied upon.  Many such tests exist, and
there are far too many to do justice to here, but see for instance the astro
code wiki\footnote{http://www-theorie.physik.unizh.ch/astrosim/code/doku.php?id=home:home},
which has a number of cross-comparison tests with other codes,
specifically aimed at disc-like models.  As the code I use is a derivative of
the one discussed in \citet{Price05} the discussion of numerical tests found
here is particularly appropriate.  A further suite of standard tests including
Sod shocks and Sedov blasts among others, used for
both code verification and comparison, is given in
\citep{Taskeretal08}.

\newpage
\bibliographystyle{Cossins}
\bibliography{Cossins}

\end{document}